\documentclass[final,5p,times,twocolumn]{elsarticle}

\usepackage[utf8]{inputenc}
\usepackage[T1]{fontenc}
\usepackage{textcomp}

\usepackage[table]{xcolor}   
\usepackage{colortbl}        
\usepackage{graphicx}
\usepackage{booktabs}
\usepackage[hidelinks]{hyperref}
\usepackage{longtable}
\usepackage{multirow}
\usepackage[super]{nth}
\usepackage{enumitem}
\usepackage{amsmath,amssymb,amsfonts}
\usepackage{caption}
\usepackage{subcaption}
\usepackage{placeins}
\usepackage{scalerel}
\usepackage{siunitx}
\usepackage{adjustbox}
\usepackage{makecell}
\usepackage{array}
\usepackage{rotating}
\usepackage{pifont}
\usepackage{algorithm}
\usepackage{algpseudocode}
\usepackage{tikz}
\usetikzlibrary{shapes.geometric, arrows, positioning, svg.path, calc}
\usepackage{orcidlink}

\definecolor{Gray}{gray}{0.9}
\definecolor{mygray}{gray}{0.9}
\newcommand{\cmark}{\ding{51}}
\newcommand{\xmark}{\ding{55}}

\DeclareUnicodeCharacter{03B8}{\ensuremath{\theta}}

\begin{document}

\title{Data Encryption Battlefield: A Deep Dive into the Dynamic Confrontations in Ransomware Attacks}

\author[csu]{Arash Mahboubi\corref{cor1}\orcidlink{0000-0002-0487-0615}}
\ead{amahboubi@csu.edu.au}

\author[unsw]{Hamed Aboutorab\orcidlink{0000-0002-9285-9917}}
\author[data61]{Seyit Camtepe\orcidlink{0000-0001-6353-8359}}
\author[unsw]{Hang Thanh Bui\orcidlink{0000-0001-6851-7717}}
\author[qut]{Khanh Luong\orcidlink{0000-0001-6981-7367}}
\author[murdoch]{Keyvan Ansari\orcidlink{0000-0002-9969-7682}}
\author[data61]{Shenlu Wang\orcidlink{0000-0002-0009-4057}}
\author[nsw]{Bazara Barry}

\cortext[cor1]{Corresponding author}

\address[csu]{Charles Sturt University, New South Wales, 2444, Australia}
\address[unsw]{University of New South Wales, Canberra, 2600, Australia}
\address[qut]{Queensland University of Technology, Queensland, 4000, Australia}
\address[data61]{CSIRO Data 61, Sydney, New South Wales, 2000, Australia}
\address[murdoch]{Murdoch University, Western Australia, 6150, Australia}
\address[nsw]{Cyber Security NSW – NSW Department of Customer Service}

\begin{abstract}
In the rapidly evolving domain of cybersecurity threats, ransomware stands out as a formidable challenge. Adversaries are increasingly employing advanced encryption techniques, such as entropy reduction using Base64 encoding, along with partial and intermittent encryption, to bypass traditional security measures and maximize their illicit gains. This study delves into the nuanced battleground between these adversaries, who are adept at refining encryption strategies to evade detection, and the defenders, who are constantly developing sophisticated countermeasures to safeguard vulnerable data assets. At the heart of our investigation is the application of online incremental machine learning algorithms designed to predict file encryption activities, even in the face of evolving adversaries' complex obfuscation tactics. Our research is underpinned by an extensive dataset, which encompasses 32.6 GB of data across 11,928 distinct files, including Microsoft Word documents (.doc), PowerPoint presentations (.ppt), Excel spreadsheets (.xlsx), and various image formats (.jpg, .jpeg, .png, .tif, .gif), PDF files (.pdf), audio files (.mp3), and video files (.mp4), all encrypted by a wide variety of 75 ransomware families. This dataset facilitates a comprehensive empirical analysis, enabling the assessment of various machine learning classifiers' effectiveness in predicting encryption events amid a range of adversarial strategies. The study's results highlight the exceptional performance of the Hoeffding Tree algorithm, which stands out for its incremental learning capabilities, making it particularly adept at identifying conventional and AES-Base64 (e.g., encryption-encoding used to reduce the entropy values) adversarial methods . In contrast, the Random Forest classifier, augmented with a warm-start feature, proves to be highly effective against the more elusive intermittent encryption techniques, underscoring the significance of bespoke machine learning solutions in navigating the dynamic and sophisticated landscape of ransomware threats.

\end{abstract}


\maketitle

\section{Introduction}

In the ever-evolving landscape of cybersecurity, the battle against ransomware represents one of the most challenging frontiers \cite{palmer2022ransomware}. This digital plague, characterized by its ability to encrypt victims' files and demand ransom for their release, has become a significant threat to individuals, organizations, and even national security. The dynamic nature of this threat is underscored by the cat-and-mouse game between cyber adversaries, notably ransomware developers, and defenders. This confrontation is emblematic of an asymmetrical battlefield where attackers, needing only to find a single vulnerability, consistently remain one step ahead of defenders who must secure every possible point of entry.

Ransomware developers are in a constant state of innovation, refining their tactics, techniques, and procedures (TTPs) to exploit the inevitable delay in defensive responses to new threats. This relentless advancement ensures their strategies are not just current but often pioneering, placing defenders in a perpetual state of catch-up. The inherent challenge in cybersecurity defense is the difficulty of anticipating and defending against the unknown. As adversaries introduce new and more complex ransomware variants, defenders are forced into a reactive stance, struggling to react effectively post-incident. For example, NetWalker, also known as Mailto, is a ransomware targeting Windows systems, first seen in 2019 and impacting healthcare, education, and government sectors. Operating as a ransomware-as-a-service (RaaS), it allows affiliates to execute attacks and share ransom profits. Its fileless nature, the fileless malware like NetWalker, however, operates directly within the computer's memory, leveraging legitimate system tools and processes to perform its malevolent actions, including file encryption and data theft. This stealthy approach not only enhances its evasion capabilities but also complicates efforts to mitigate and eradicate the ransomware from infected systems \cite{8538400}.

The heart of this challenge lies in the limitations of current ransomware detection methodologies, including machine learning models, static analysis, user behavior analytics, and dynamic analysis. Each of these approaches, while comprehensive in their scope, often lags behind the sophisticated and continuously evolving offensive tactics employed by ransomware developers. For instance, machine learning models, which are trained on known malware samples, find it difficult to identify zero-day attacks that display previously unseen behaviors \cite{WANG2023200280}. Static analysis tools, on the other hand, are increasingly evaded by ransomware that employs sophisticated obfuscation techniques. User behavior analytics can result in high false positive rates due to the variability in legitimate user activities, and dynamic analysis might be circumvented by ransomware that can detect and alter its behavior in sandboxed environments.

The future of ransomware defense is poised at a critical juncture, necessitating a departure from traditional detection methodologies towards the adoption of more predictive and adaptive strategies. Our research endeavors to explore the sophisticated tactics employed by adversaries to undermine the efficacy of existing defense mechanisms, particularly those operating at the file system level. These adversaries skillfully navigate around conventional security measures by employing strategies such as entropy value reduction, intermittent, and partial encryption, thereby diminishing the impact and detection capabilities of current defense systems.

Furthermore, our study delves into the potential of online incremental learning as a pivotal technology to enhance the differentiation between encrypted and normal files within hot data storage environments \cite{sabbaghi2017hybrid}. This approach is particularly relevant in the context of ransomware attacks that utilize intermittent and partial encryption techniques to evade detection. Our research aims to continuously update the detection models with new data (encrypted and normal), enabling them to adapt to evolving ransomware encryption tactics dynamically.

Our focus lies on the identification of unique file attributes, represented as feature vectors, that serve as effective countermeasures against adversarial actions. Several related works rely on entropy-based detection \cite{9541344, may2019combating, e24020239, McIntosh}, involve frequent file system operations \cite{Continella} and system logs frequent pattern mining \cite{8051108}. These examples of mitigation strategies, used by defenders, can be evaded using tactics such as intermittent encryption, partial encryption, and memory mapping. These tactics are discussed in detail in Section \ref{sec: Add_game}. We place a spotlight on features revealing the diversity of file types within systems, particularly under the online incremental machine learning framework \cite{8759919}. The contributions of this paper are manifold and can be delineated as follows:

\begin{enumerate}
\item We formalize existing adversarial encryption techniques used by ransomware developers, highlighting their capability to circumvent traditional security measures.
\item We identify unique feature vectors capable of distinguishing between encrypted and unencrypted data at the file system level.
\item Investigating ML models that can reduce the computational costs associated with the need for CPU- and memory-intensive tasks.
\item Through empirical research, we investigate strategic approaches for both attackers, aiming to minimize entropy to evade detection, and defenders, using entropy as a pivotal metric for spotting encrypted data. Our study particularly examines the impact of Base64 encoding on entropy reduction and evaluates its potential exploitation by adversaries to lower data entropy successfully. 
\item We evaluate the efficacy of machine learning models in real-world ransomware scenarios, aiming to thwart ransomware attacks during their encryption phase. This includes determining alert activation thresholds and developing adaptive strategies to respond to evolving data patterns.
\end{enumerate}

The structure of this paper is organized as follows: In Section \ref{sec: Threat_mdodel}, we outline the threat model and foundational assumptions underpinning this study. Section \ref{sec: Add_game} outlines methodologies to understand strategies in adversarial and defensive scenarios, focusing on advanced encryption techniques used by ransomware developers. Moving on to Section \ref{sec: Methdology}, we describe our methodology and approach. In Section \ref{sec: Evaluation&Result}, we conduct a comprehensive analysis of machine learning paradigms, with a specific focus on both shallow and deep learning classifiers. To evaluate their efficacy, we perform targeted micro-experiments. Additionally, we articulate the core principles associated with online incremental machine learning, tailored for the identification of encrypted files within a file-level system. Subsequently, Section \ref{Sec: Related_Work} reviews existing literature and elucidates the principles that inform it. Finally, Section \ref{sec: Conclusion} offers concluding remarks and outlines prospective directions for future research.

\section{Threat model}\label{sec: Threat_mdodel}
In our threat model, we presume that adversaries possess comprehensive knowledge of the proposed system and the features extracted from the files. This is grounded in the practical realities of the cybersecurity landscape, where sophisticated adversaries are often privy to significant information about the systems they target. For our experiments, we employ real-world ransomware to encrypt files, replicating the tactics commonly deployed by adversaries, such as partial encryption and intermittent encryption.

It should be clearly stated that our systematization of knowledge investigation does not aim to detect the ransomware binary or its behavior patterns. Rather, our main objective is to ascertain the encryption status of a multitude of file types within a file system. By focusing our efforts on this aspect, we aim to develop a robust and accurate tool that can contribute to the mitigation of damage and data loss from ransomware attacks, rather than identifying the ransomware itself. This approach serves as a complement to other strategies and models focusing on ransomware detection and behavior analysis.

To understand countermeasures and adversaries' techniques, we elucidate the interplay between adversaries and defenders using the framework of strategic games. By framing the cybersecurity landscape as a strategic game, we can formally analyze both defenders' and adversaries' system vulnerabilities and weak points, particularly during the initial stages of ransomware file encryption. Viewing this through the lens of strategic games allows us to evaluate decision-making processes, simulate possible responses, and predict potential outcomes.

We operate under the assumption that online learning is particularly suitable for data streams. After conducting a systematic exploration of adversarial tactics and defensive strategies, our objective is to incorporate machine learning, specifically online learning algorithms, into backup systems and network file-sharing drives, including, but not limited to, NFS and SMB protocols. The model serves the purpose of detecting unauthorized encryption activities before the initiation of any protocol-based encryption processes. In the realm of file-level systems, this integration can be achieved through a variety of open-source file systems capable of embedding the proposed online incremental machine learning model. Prominent examples are FUSE (File system in Userspace) and Puffs (Pass-to-Userspace Framework File System), which are elaborated upon in \cite{Mahboubi2022} and \cite{267682}, respectively.

\section{Adversarial strategies} \label{sec: Add_game}
In this section, we delineate formal methodologies for comprehensively understanding the strategies employed in both adversarial and defensive scenarios, with a particular focus on sophisticated encryption techniques devised by ransomware developers.

\subsection{Legitimate encryption process usage}
The complexity of the game is accentuated by the attacker's strategy of co-opting legitimate system processes to carry out malicious activities. This deceptive approach, commonly termed a 'stealth attack,' obfuscates the distinction between benign and malicious operations, thereby posing a significant challenge for the defender in neutralizing the threat without adversely affecting regular system functions. 

\begin{itemize}
    \item  Let us denote $( E )$ as the set of legitimate encryption processes that can be used by the operating system to encrypt files and $( R )$ as the ransomware process. Additionally, let $( A )$ be the adversary or ransomware attacker, $( D )$ as the system defender, and $( P )$ as the set of all system processes. Furthermore, $( M )$ is a mapping function that maps a process to its legitimacy status, such that $( M : P \rightarrow {0,1 } $), where `$0$' indicates malicious and `$1$' indicates benign.    
    
    In a typical scenario, a system defender $( D )$ tries to neutralize malicious processes, i.e., for any process $( p )$ in $( P )$, if $( M(p) = 0 )$, then neutralize $( p )$. However, the complexity arises when ransomware $( R )$ employs legitimate processes $( E )$ for encryption, i.e., $( R(E) )$. In this case, $( M(R(E)) = 1 )$, making it seem benign to the system defender $( D )$.
    
    Given this, it becomes a challenge for the defender to distinguish between benign and malicious activities, as the ransomware attacker $( A )$ has essentially turned the scenario into a game where $( A )$ and $( D )$ have conflicting objectives. While $( A )$ tries to maximize the use of legitimate processes $( E )$ for malicious deeds, $( D )$ attempts to neutralize malicious processes without disrupting regular operations.
    
    Therefore, there is a need for a more nuanced approach that can distinguish between the legitimate use of processes \( E \) and their malicious use by \( R \), possibly by considering additional context or behavioral patterns. However, this task is non-trivial and represents a significant challenge in the current landscape of ransomware detection and neutralization.
\end{itemize}

\subsection{Gradual write I/O to storage}

Typically, defenders scrutinize file system operations, including reading, writing, deletion, and renaming, by utilizing predetermined threshold values as a means to detect ransomware activity. Nevertheless, ransomware employs a strategy that resorts to memory mapping to circumvent the file I/O read and write behaviors. In such circumstances, the existing countermeasures applied by defenders are prone to failure.

Memory mapping is a technique that allows programs to interact with data in storage as if it were in the computer's main memory. The operating system creates a mapping between the program's address space and the storage, within the context of virtual memory. This enables the program to use standard memory access instructions for file operations, potentially simplifying and improving efficiency over traditional file access methods.

\begin{itemize}
    \item Let $( O )$ be the set of file operations including reading, writing, deletion, and renaming, and $( T )$ be the predetermined threshold values used to detect ransomware activity. Traditional detection methods focus on monitoring these operations such that if any operation $( o )$ in $( O )$ exceeds the corresponding threshold $( t )$ in $( T )$, an alarm is raised. Formally, for any operation $( o )$ in $( O )$, if $( o > t )$, then an alarm is raised.

    However, ransomware employs memory mapping which is a process by which the operating system creates a mapping $( M )$ between the address space of the process $( P )$ (the range of addresses that the process can use to address memory) and the storage object $( S )$. This can be represented as $( M : P \rightarrow S )$. When this mapping is in place, file I/O operations are transformed into memory access operations.
    
    Consequently, if ransomware $( R )$ utilizes memory mapping, the file I/O operations become memory operations and can evade detection by systems strictly monitoring file access. So for any file operation $( o )$ in $( O )$ performed by ransomware $( R )$, if $( R )$ employs memory mapping, $( o )$ is transformed into memory operation $( m )$, i.e., $( o \rightarrow m )$. Since defenders are monitoring $( O )$ and not $( m )$, these operations can potentially bypass detection.
  
\end{itemize}

There are number of research studies in \cite{AhmedMuhammad, 1Continella,  HIRANO2022301314, 8939214, Huang, jeong2019anomaly, 197235, Mehnaz, Milajerdi, 9493745,7536529} using monitoring methods often focus on file system changes, and when ransomware uses memory mapping, these file I/O operations are essentially translated into memory operations. This means that changes that would ordinarily be tracked (like file creation, deletion, renaming, and modifications) may not be detected, leading to a potential blind spot in ransomware detection mechanisms.

\subsection{Partial data encryption} 
Ransomware strategies, distinguished by their swift and elusive encryption techniques, exhibit intriguing characteristics, such as the adaptable partial encryption of files exceeding 5.245 MB \cite{constantin2022new}. Although partial file encryption is not an innovative tactic - various ransomware programs deploy this to accelerate the process - the novel feature lies in its ability to specify the quantum of a file to encrypt. This capability bears significant implications for security programs that traditionally monitor file modifications to identify potential ransomware incursions. The ensuing fragmentation and probable low percentage of encrypted file content reduces the likelihood of detection by defenders. 

This encryption methodology, along with other tactics employed by Royal ransomware, bears resemblance to Conti ransomware. For instance, the Conti ransomware also utilized 5.24MB as a threshold for partial encryption, subsequently segmenting the file into several equal parts, encrypting one part and leaving the next unencrypted. However, Conti's approach diverges by encrypting 50\% of those segments. This partial encryption can be dynamic. Therefore, defensive strategies may fail if they rely on changes in file system attributes or CPU arithmetic operations like the Exclusive OR (XOR) logic gate. 

The several potential modes of partial encryption used by real-world ransomware include:

\begin{itemize}
    \item \textbf{Skip-step encryption mode:} Let \( F \) denote the file to be encrypted, where \( F_i \) represents the \( i^{th} \) megabyte (MB) of the file \( F \). Let \( N \) and \( Y \) be non-negative integers. Then, in the skip-step mode, the file \( F \) is encrypted such that:
    \[
    \forall i, F_i \text{ is encrypted } \iff i \mod (N+Y) \geq N
    \]
    This represents an encryption process that skips the first \( N \) MB, then encrypts the next \( Y \) MB, and so on.
\end{itemize}

\begin{itemize}
    \item \textbf{Fast encryption mode:}
    Let \( F \) and \( F_i \) denote the same as above, and \( N \) be a non-negative integer. Then, in the fast mode, the file \( F \) is encrypted such that:
    \[
    \forall i, F_i \text{ is encrypted } \iff i < N
    \]
    This indicates an encryption process that encrypts only the first \( N \) MB of the file \( F \).
\end{itemize}

\begin{itemize}
    \item \textbf{Percent encryption mode:}
    Let $ ( F )$ and $( F_i )$ denote the same as above, and $( N )$ be a non-negative integer, and $( P )$ be a percentage $(0 \leq P \leq 100)$. 
    Also, let $( S )$ be the size of the file in MB, and $( P_{MB} = \frac{P}{100} \times S )$.
    Then, in the percent mode, the file $( F )$ is encrypted such that:
    \[
    \forall i, F_i \text{ is encrypted } \iff i \mod (N + P_{MB}) \geq P_{MB}
    \]
    This represents an encryption process that skips $( P_{MB} )$ (which equals $( P\% )$ of the total file size), then encrypts every $( N )$ MB, and so on.

\end{itemize}

\subsection{Intermittent encryption}

Intermittent encryption strategies involve selectively encrypting data based on specific criteria. This approach allows for dynamic encryption, optimizing security and efficiency by adapting to varying contexts.

The intermittent encryption strategy \cite{toulas2022ransomware}, as used by LockFile and Black Basta ransomware, significantly differs from the partial encryption techniques adopted by LockBit 2.0, DarkSide, and BlackMatter. While partial encryption, targeting the initial segments of documents, aims to expedite the process, intermittent encryption focuses on encrypting alternate 16-byte segments of a file. This results in scrambled and untouched data alternating throughout the file.

Although this strategy is slower than partial encryption, it disrupts statistical analysis which is a key tool in ransomware detection. Detection programs typically block any process from modifying additional files if the statistical analysis test indicates encryption. With encrypted files appearing significantly different from unencrypted ones in statistical analysis, a clear difference can be seen in the chi-squared ($chi^2$) test scores or entropy value for files encrypted by ransomware.

The Intermittent Encryption strategy used by ransomware brings an additional layer of complexity for defenders where the ransomware encrypts pieces of a file at regular intervals. Ransomware can use the option to intermittently encrypt data, which is a feature that can be adjusted and customized. 

For instance, BlackCat ransomware introduces a versatile implementation of intermittent encryption, enabling operators to select from an array of byte-skipping patterns. This flexibility allows for diverse encryption modes that can enhance the complexity and potentially the security of the encrypted data, such as:

\begin{itemize}
    \item \textbf{SmartPattern [N,P]:}
    Encrypts N megabytes of the file in percentage steps. For example, as its default setting it starts from the file's header and encrypts 10 megabytes every 10\%.

\end{itemize}

\begin{itemize}
    \item \textbf{Auto mode:}
    This mode amalgamates multiple encryption methods for a more convoluted result. The encryption pattern in this mode is a complex function of the different encryption methods available, potentially including all the previously mentioned modes and their parameters.

\end{itemize}

Another intermittent encryption is Black Basta ransomware, a prominent figure in the cybercrime space that operates differently. Its strain of ransomware makes decisions based on the size of the file rather than offering operator-selected modes.

\begin{itemize}
    \item \textbf{For small files:}
    For a file \( F \) of size \( S \), \( F \) is encrypted if \( S < 704 \) bytes. 
\end{itemize}

\begin{itemize}
    \item \textbf{For medium size files:}
    For a file \( F \) with \( S \) in the range 704 bytes \( \leq S < 4 \) KB, for every \( b^{th} \) byte \( F_b \), \( F_b \) is encrypted if \( b \mod 256 < 64 \).
\end{itemize}

\begin{itemize}
    \item \textbf{For larger files:}
    For a file \( F \) with \( S > 4 \) KB, for every \( b^{th} \) byte \( F_b \), \( F_b \) is encrypted if \( b \mod 192 < 64 \).
\end{itemize}

Notably, if the defender strategy is calibrated to react only to significant statistical differences to prevent false positives, it could fail to detect the encryption performed by ransomware. Additionally, the ransomware strategy complicates incident response efforts by deleting itself after completing the encryption process, making it difficult for defenders to locate a ransomware binary for analysis and system cleansing.

\subsection{Adversaries avoidance of decoy techniques} In the ongoing struggle against nefarious file system activities, both commercial and research anti-ransomware strategies have adopted the use of deception-based methodologies, particularly the strategic placement of decoy files amongst authentic user files. This approach introduces an additional layer of complexity to ransomware detection efforts, predicated on the understanding that any interaction with a decoy file inherently signifies a malicious activity. However, in order to illustrate how ransomware can effectively circumvent existing deception-based detection strategies, researchers have put forth a proof-of-concept for anti-decoy ransomware in a scholarly publication \cite{ZiyaAlper}. This lab-developed ransomware is equipped with a decision engine that employs a minimal set of rules, thereby successfully evading decoys. Here an abstract analysis of the approach to bypass the decoy strategy can be used by an adversary. 

\begin{itemize}
    \item \textbf{Detecting static decoys through heuristics:} This strategy involves ransomware using heuristics to identify patterns that suggest the presence of decoy files. These patterns could be files filled with empty values or those with static creation dates and content. By fingerprinting these checks at run-time, ransomware can exclude these decoy files from the target files to encrypt. It notes that if ransomware mistakenly identifies a user file as a decoy and excludes it, the impact on the overall strategy is minimal as the ransomware will still encrypt other files. Two heuristic methods are suggested; one targets hidden and empty files, while the other aims at non-regular files like symbolic links or named pipes.
    \item \textbf{Distinguishing decoys using statistical methods:} This method is based on understanding file storage on the Windows operating system to discern file attributes and metadata. The paper mentions that decoy files, which are not typically accessed by users, will show different access patterns than genuine files. By analyzing these differences statistically, it's possible to distinguish between decoy and genuine files. The technique uses 36 metrics for each directory based on file attributes and their standard deviations, creating a feature vector of 72 metrics. This statistical method seems effective at finding discrepancies, especially with time and date information.
    \item \textbf{Monitoring User to Reveal Non-decoy Files:} This technique focuses on monitoring user activities to identify genuine files. Two strategies are suggested. The first involves injecting a spy module into Explorer.exe to monitor which files are accessed by user applications. The second proposes enumerating all processes and injecting an interceptor module, replacing the WriteFile API with encryption routines.

\end{itemize}

\subsection{File entropy and data manipulation strategy}

The strategy employed by defenders to monitor potential shifts in file entropy, which might serve as indicators of ransomware's cryptographic activities, has been scrutinized and studied \cite{9541344, 9493745, KharrazAmin, 7536529}. In one study \cite{McIntosh}, it was claimed that no instance of ransomware was discovered that manipulates the entropy values of encrypted content. Nevertheless, we have highlighted certain ransomware families that utilize techniques such as partial encryption and intermittent encryption. Moreover, Maze ransomware \cite{Mahboubi2022}, is known for its insertion of "Null" characters as a strategy to reduce the entropy of an encrypted file prior to its storage.

The referenced study \cite{McIntosh} evaluates the entropy-based changes experienced by diverse file types as a result of Base64-Encoding and either partial or full encryption. Our extensive analysis suggests the existence of at least three methodologies by which file entropy values can be manipulated. This supposition adds another layer to the complex landscape of ransomware evasion tactics and challenges the robustness of entropy-based detection strategies. Therefore, an attacker's strategy might involve reducing the file entropy values after the  encryption and encoding phases, while the defender could rely on measuring the uncertainty, unpredictability, or randomness of the file at the storage level.

\subsubsection{Base64 encoding}

Various studies have suggested approaches for countering ransomware detection by using a Base64 encoding algorithm to lower file entropy \cite{e24020239, McIntosh}. The Base64 algorithm converts binary data into ASCII text by transforming it into a radix-64 format \cite{kumbhojkar2021base64}.

The objective of Base64 encoding is to transform a binary data stream \( D \) comprised of \( n \) 8-bit bytes \( b_1, b_2, \ldots, b_n \) into an ASCII string \( S \). The algorithm partitions the \( n \) bytes into blocks of three, concatenates each block to form a 24-bit block, and divides this 24-bit block into four 6-bit segments. Each 6-bit segment is then mapped to an ASCII character from the Base64 alphabet, usually composed of 64 printable ASCII characters ranging from 'A' to 'Z', 'a' to 'z', '0' to '9', '+', and '/'. Padding is applied when the last block contains fewer than 3 bytes, filling with zero bits and appending one or two '=' characters as needed. 

Mathematically, for a data stream \( D = b_1b_2 \ldots b_n \), it is partitioned into \( k \) blocks \( D_1, D_2, \ldots, D_k \) where \( D_i = (b_{3i-2}, b_{3i-1}, b_{3i}) \). Each \( D_i \) produces a 24-bit block \( B_i = b_{3i-2} \parallel b_{3i-1} \parallel b_{3i} \), which is subdivided into four 6-bit segments \( S_{i1}, S_{i2}, S_{i3}, S_{i4} \). These segments are mapped to ASCII characters \( C_{ij} = \text{Base64}(S_{ij}) \). The output \( S \) is the concatenation \( C_{11}C_{12}C_{13}C_{14}C_{21}C_{22} \ldots C_{k4} \). This facilitates the safe transfer of binary data over systems optimized for text-based data.

\begin{figure} [ht!]
    \centering
    \includegraphics[width=8.5cm]{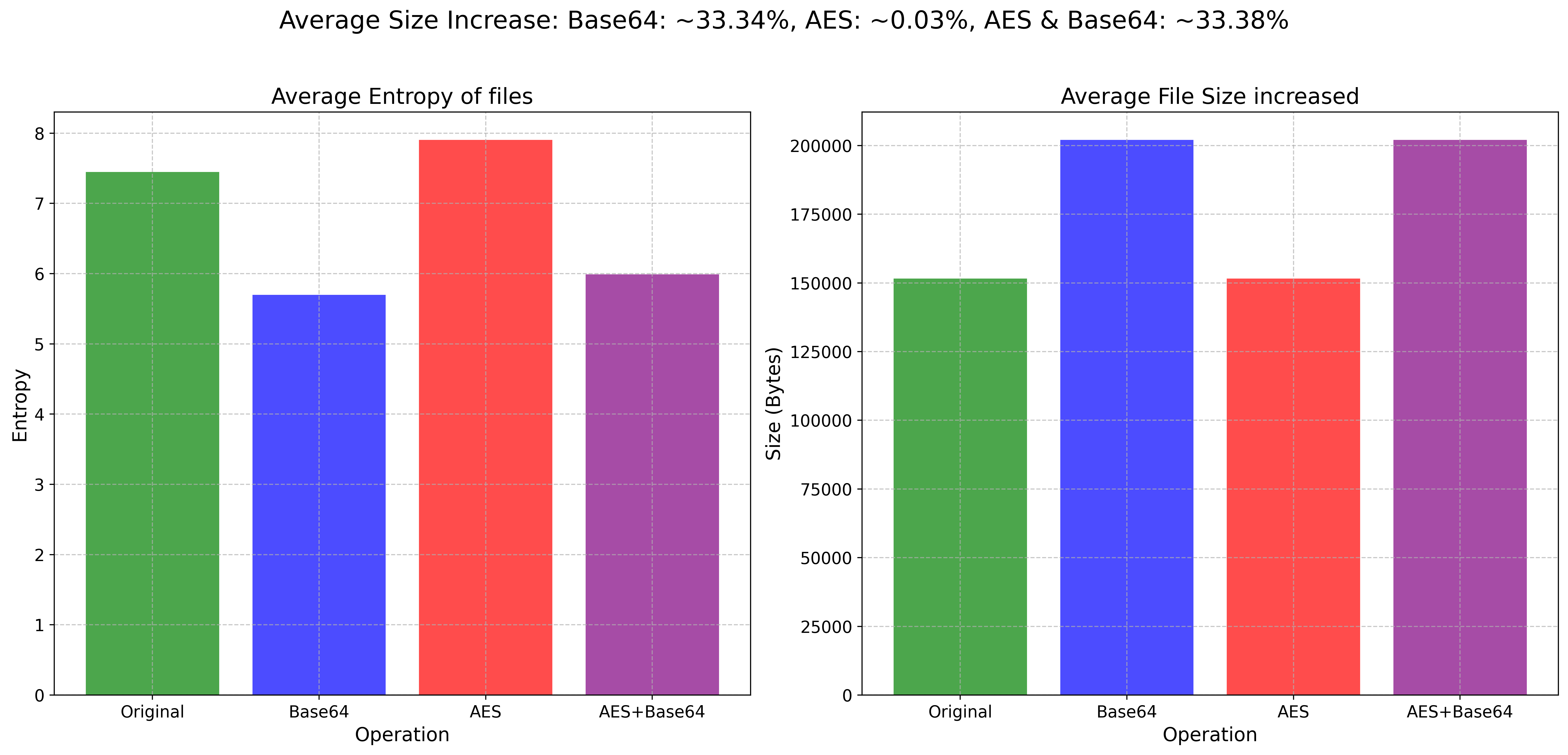}
    \caption{ Base64 encoding reduces the entropy from 7.99 to 5.99. Adversary seeking to minimize entropy, a sequential AES $\xrightarrow[]{}$ Base64 encoding approach is a tangible approach. However, this results in a $~33.38\%$ increase in the original file size.}
    \label{fig:base64}
\end{figure}

We conduct an experiment to determine the most effective strategy for an adversary aiming to reduce data entropy. But, first, it is well-known that Base64 encoding increases the size of the data by approximately 33\% \cite{microsoft2023configure}. Our focus is on a sequential data transformation process aimed at entropy reduction. Initially, we gathered raw binary data from a dataset consisting of 3,200 files (8.13 GB) across multiple formats, including JPG, PDF, Microsoft documents, and TIFF. The entropy of this raw data is calculated using Shannon's entropy formula to establish a baseline. The data then undergoes a first layer of transformation via Base64 encoding, which serves to standardize the data representation and generally aims to reduce its entropy. This stage is represented by the blue bar in Figure \ref{fig:base64}. Subsequently, the data is encrypted using AES, a step that significantly increases its entropy. This transformation is depicted by the red bar in Figure \ref{fig:base64}. Finally, we apply a combination of AES encryption and Base64 encoding. This multi-step process successfully reduces the entropy of the files, as illustrated by the purple bar in Figure \ref{fig:base64}. However, this results in a $~33.38\%$ increase in the original file size. 

\begin{figure} [ht!]
    \centering
    \includegraphics[width=8.3cm]{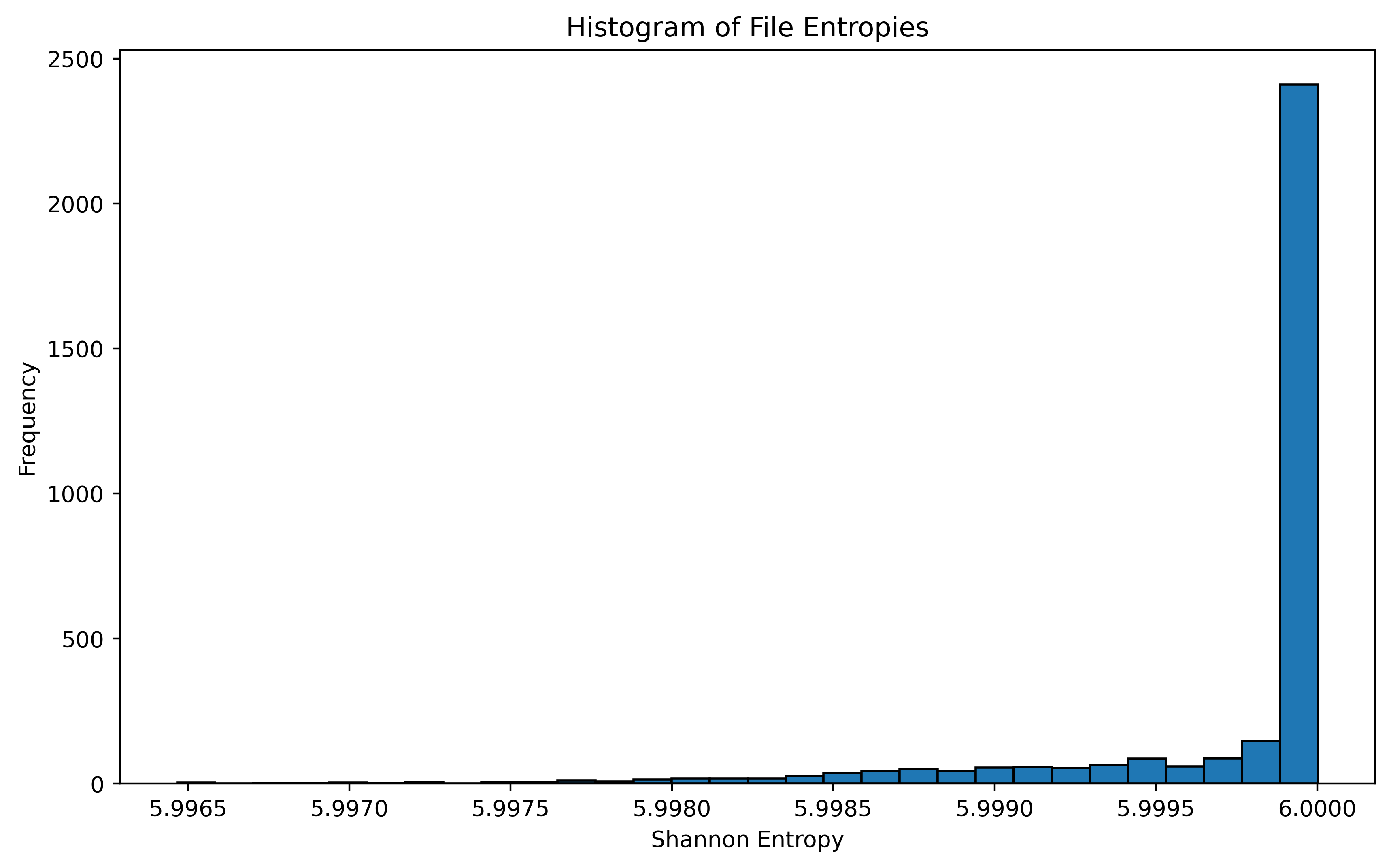}
    \caption{ Base64 encoding reduces the entropy from 7.99 to 5.99. However, in a sequential AES $\xrightarrow[]{}$ Base64 encoding approach, the Shannon entropy results fall in the range of 5.99 to 6.} 
    \label{fig:Shannon entropy}
\end{figure}

In our empirical study, we observed that the Shannon entropy values for data processed through an Encrypt-Base64 encoding pipeline consistently fall within the range of 5.99 to 6.0. These findings imply that defensive mechanisms could leverage this specific range of entropy values to improve the detection of encrypted data stores. Figure \ref{fig:Shannon entropy} presents the Shannon entropy measurements across 3200 different types of files. But the question is, will adversaries risk incorporating Base64 encoding into their strategy, despite being aware that an increase in file size may potentially trigger other defense systems?

\subsection{Adversary falsifies the defender's strategy by maintaining a uniform distribution of symbols} \label{sec: Optimal defense}

Many proposed models rely on measuring of file entropy for detecting ransomware \cite{e24101503, 8772046, DAVIES2021102377, e24020239, 7536529, kharaz2016unveil, Mehnaz}. Entropy in information theory facilitates the establishment of a uniform distribution of symbols, an essential precondition for ensuring high encryption security. Under ideal circumstances, every possible outcome within a uniform distribution should hold an equivalent likelihood, thereby fostering maximum unpredictability. If an encrypted file upholds this uniform distribution, it indicates successful randomization, making it challenging for cryptoanalysis to detect patterns or infer original data. 

However, failure to maintain a uniform distribution might suggest the existence of identifiable patterns in the encryption. These patterns could be manipulated by either a defender or an adversary, leading to decreased security. This is particularly critical when an adversary is attempting to implement robust encryption while preserving the ability to decrypt, highlighting an inherent paradox within the ransomware business model.

On the one hand, the attacker must implement robust encryption to prevent access to the victim's data. Conversely, they must also retain the ability to reliably reverse this encryption when necessary (if and only if the victim pays the ransom). An inability to accomplish this dual objective threatens the viability of the ransomware business model, removing the victim's incentive to pay the ransom.

An adversary's strategic move in this confrontation often involves the use of partial or intermittent encryption, a tactic aiming to lower the entropy value while maintaining the legitimacy of the ransomware business model. This strategy's successful execution could signify a potential loss for the defender in this strategic standoff.


\section{Methodology} \label{sec: Methdology}

In our research, we meticulously aim to evaluate the compatibility and efficacy of various machine learning models, with a primary focus on file system data encryption realms. Our methodological framework is robustly constructed, guiding the identification and validation of optimal machine learning models. These models are anticipated to make substantial contributions to future works aimed at detecting and mitigating the challenges posed by the cryptographic landscapes of ransomware.

\paragraph{\textbf{Phase 1: Baseline Analysis with Traditional Machine Learning Models}}
Initially, we assess a range of popular non-online machine learning classifiers to understand their efficacy in differentiating between encrypted and unencrypted data buffers at the storage level \cite{MAHBOUBI2024103873}. The classifiers chosen for this phase represent a diverse set of well-established machine learning paradigms, as detailed in previous literature \cite{alzubi2018machine, das2017survey}. Our primary goal here is to establish a baseline and comprehend the limitations of traditional machine learning techniques in real-time classification tasks.

\paragraph{\textbf{Phase 2: Hypothesis Formation and Real-time Use Cases}}
Drawing from the foundational insights garnered in Phase 1, we have developed a hypothesis that online learning methods could be advantageous in real-time detection scenarios, specifically in identifying ransomware attacks (i.e., file encryption). We are exploring the possibility of deploying classifiers, based on online learning, in dynamic environments such as large corporate offices or multi-tenant cloud storage systems. 

\paragraph{\textbf{Phase 3: Selection of Online Learning Models}}
Subsequently, we narrow our focus to first-order online learning models, which promise the real-time adaptation essential for dynamic environments. For a more in-depth investigation, we select three leading online learning algorithms: Stochastic Gradient Descent (SGD), Perceptron, and Passive-Aggressive algorithms \cite{HOI2021249}. The choice of these models is not arbitrary but is influenced by the limitations we identified in traditional machine learning classifiers. To refine our model selection further, we also explore other compatible linear models and ensemble methods for online learning \cite{alzubi2018machine, das2017survey}.

\paragraph{\textbf{Phase 4: Addressing the Challenges with Stream Data}}
We recognize that both traditional and online learning algorithms may still fall short in capturing the complexities of stream data, which is often encountered in the scenarios we consider. The dynamic nature of these scenarios—ranging from ransomware attacks to routine organizational data flow-calls for specialized algorithms designed to handle continuous data streams.

\paragraph{\textbf{Phase 5: Final Model Selection}}
Given the challenges outlined in Phase 4, the Hoeffding algorithm is employed for our study. This decision tree learning method is specifically designed for classifying stream data. It provides a robust mechanism for real-time classification of encrypted and unencrypted data, thereby aligning with our objective for immediate detection and response in various scenarios. 


\begin{figure}
    \centering
    \includegraphics[width=6cm, height=7cm]{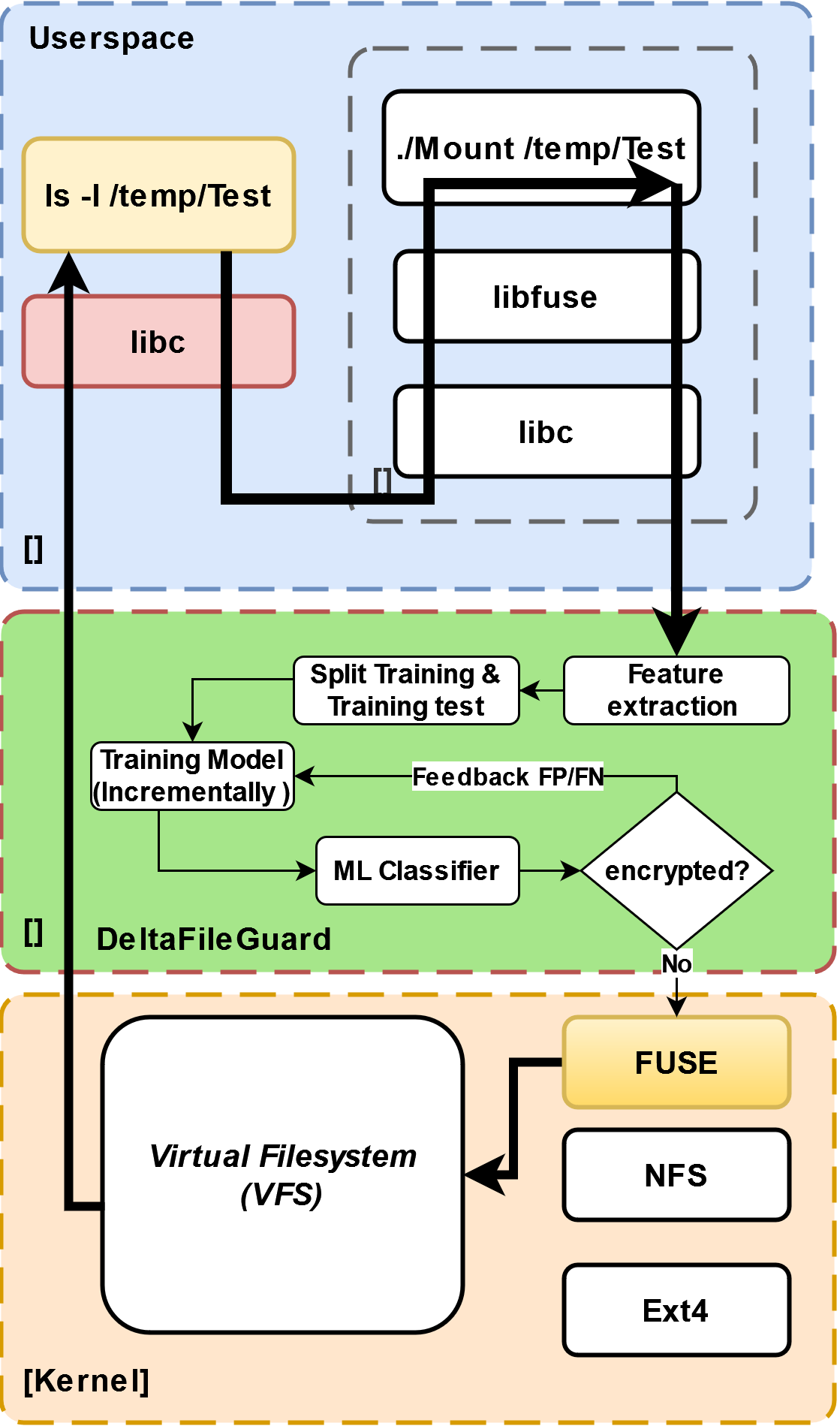}
    \caption{Conceptualize diagram: DeltaFile Guard represents the utilization of online incremental learning within the FUSE file system.}
    \label{fig:DeltaFile}
\end{figure}

\subsection*{\textbf{Approach to handling file operations and the core components}}

Integrating DeltaFile Guard with the FUSE file system enables a sophisticated approach to handling file operations with an added layer of intelligence. We have developed a Middle Layer component to accommodate online incremental learning illustrated in Figure \ref{fig:DeltaFile}. Here is a detailed breakdown of how this integration facilitates online incremental learning to predict the encryption status of files based on extracted features:
\begin{itemize}
    \item \textbf{DeltaFile Guard:} This represents the specialized component designed to intercept file operations within the FUSE file system. Its main role is to apply machine learning algorithms for real-time analysis and prediction. DeltaFile Guard focuses on identifying whether files are encrypted, leveraging online incremental learning to adapt its models based on incoming data.

    \item \textbf{FUSE File System:} A flexible user space file system framework that allows custom file system operations to be implemented without altering the kernel. It serves as the foundation upon which DeltaFile Guard operates, providing the necessary infrastructure for file manipulation and access.

\end{itemize}

\subsection*{\textbf{Workflow}}

As files are accessed by the write function within the FUSE file system, DeltaFile Guard intercepts these operations. This interception is crucial for analyzing file content and metadata without disrupting the user’s interaction with the file system.

\begin{itemize}
    \item \textbf{Feature Extraction:} For each file operation, DeltaFile Guard extracts relevant features. These features are discussed in the following feature engineering section. 
    
    \item \textbf{Online Incremental Learning:} Utilizing the extracted features, the embedded machine learning model within DeltaFile Guard performs real-time predictions to determine if a file is encrypted. The "online incremental" aspect refers to the model's ability to learn and update its parameters dynamically as new data arrives, ensuring the system evolves and adapts to new patterns or encryption methods over time.

    \item \textbf{Prediction and Action:} Based on the prediction outcome, DeltaFile Guard can take predefined actions. For encrypted files, it flags them for further review and trigger alerts. For non-encrypted files, it ensures they are handled per standard file system operations.
    
\end{itemize}

\section{Empirical evaluation and results} \label{sec: Evaluation&Result}

\subsection{Dataset}
In this section, we implement a structured empirical approach to appraise the performance of a multitude of machine learning classifiers. We require an appropriate classification function with the intent of augmenting the probability of detecting encrypted data whilst concurrently reducing the occurrence of false positives and false negatives.

For this study, a comprehensive set of 75 unique ransomware families was curated from publicly available malware repositories. utilizing controlled environments, each ransomware variant was executed against a dataset comprising approximately 32.6 GB, encompassing a total of 11,928 files. This dataset spans a broad array of file formats, including but not limited to mp3, mp4, docx, pptx, png, pdf, jpeg, gif, xls, and csv. The assembled dataset features a mixture of encrypted and non-encrypted files, designed to mimic a genuine computational setting. Notably, the encrypted portion of this dataset has been generated using real-world ransomware, ensuring the inclusion of the latest ransomware families in our analysis.

\subsection{Ransomware dataset} \label{App: Dataset}

In the course of our research, we amassed a comprehensive collection of ransomware samples sourced from various public repositories, including but not limited to VirusShare, MalwareBazar, MalwareDB, and the Zoo. Our ensemble consists of approximately 1,500 distinct samples. To gain insights into the characteristics and behavior of these malicious entities, we employed both static and dynamic analysis techniques. Prior to in-depth analysis, each sample underwent rigorous testing to ensure its ability to execute, thereby allowing us to accurately observe and document its inherent behavior.

\subsubsection{System specification}
Our experimental environment is powered by the AMD Ryzen ThreadRipper Pro 5975WX 32-core, paired with 32 GB of RAM and a 1TB SSD, providing ample processing power and storage capacity for handling extensive datasets and complex models. This setup operates on Windows 11 and leverages Python 3.10 and the Spyder integrated development environment (IDE) for seamless machine learning code development and experimentation.

\subsection{Feature engineering} \label{sec:feature}

We preprocess the data by extracting relevant features from the files, preprocessing data through extraction of pertinent features plays a significant role in enabling classifiers to effectively distinguish between encrypted and unencrypted files. Each selected feature offers a distinctive perspective on the characteristics of the file content, thus providing the classifier with a comprehensive understanding of its structure and composition.

Our approach seeks to augment traditional detection methods, which primarily focus on entropy and file size, with a suite of more nuanced features. Traditional methods are increasingly circumvented by advanced encryption schemes designed to evade detection, highlighting the need for a broader set of characteristics that can reveal the presence of encryption. While file size and byte entropy are foundational, they offer limited insight into sophisticated evasion techniques. Our inclusion of content-based features such as byte frequency, variance, kurtosis, skewness, and the analysis of strings and file content patterns, delves deeper into the structural intricacies of files. These features are designed to detect subtle anomalies and patterns (or their absence) indicative of encryption, thereby providing a more robust framework for distinguishing encrypted files from their unencrypted counterparts.

Moreover, our methodology introduces novel features such as entropy variance and percentiles utilization, which are not commonly employed in existing solutions. The entropy variance feature aims to capture the uniformity of randomness across a file, a characteristic trait of encrypted content, while percentiles utilization examines the distribution of byte values at specific intervals, offering insights into the uniformity expected from encrypted files versus the varied distributions of plaintext files. These novel features, particularly when combined with traditional metrics, enhance the detection of encrypted files by identifying characteristics that evade simpler detection methods. This comprehensive approach not only addresses the limitations of relying solely on entropy and file size but also sets a new standard for encrypted file detection by incorporating a more detailed analysis of file characteristics, thereby improving the ability to detect advanced encryption schemes.

\begin{enumerate}
\item \textbf{File Size:} This attribute provides a trivial understanding of a file's intricacy and potential data volume. Notably, encrypted files, contingent on the applied encryption methodology, may exhibit unique size patterns that starkly contrast those of their unencrypted counterparts. Later in the context of online learning, we opt to exclude this feature since the file size may be subject to constant changes.

\item \textbf{Byte Entropy:} This feature, indicative of a file's randomness or unpredictability, generally presents higher values in encrypted files. The calculation of byte entropy serves as a tool for the identification of encrypted files due to their inherently unpredictable nature.

\begin{align}
H(X) & = -\sum_{i=1}^{n} p(x_i) \log_b p(x_i)
\end{align}

Where $X$ is a random discrete variable, $x_i$ is a possible value of $X$, $p(x_i)$ is the probability mass function of $X$ at $x_i$, and $b$ is the base of the logarithm.

\item \textbf{Content-Based Features:} Including the frequency or scarcity of specific characters or sequences within file content, these features aid in unveiling patterns or consistencies present in unencrypted files but absent in encrypted ones.

\paragraph{Byte frequency:}

\begin{align}
f_i & = \frac{\text{count}(x_i)}{N}
\end{align}

In Byte frequency, $x_i$ is a byte value and $N$ is the total number of bytes.

\item \textbf{Byte Variance:} As a measure of dispersion, byte variance can offer valuable insights into the range of byte values in a file. The high variance may suggest a broader range of byte values, a characteristic potentially indicative of encrypted files. 

\begin{align}
   \sigma^2 & = \frac{1}{N}\sum_{i=1}^{N}(x_i-\mu)^2
\end{align}

\item \textbf{Byte Kurtosis:} This measure of data point distribution can provide a statistical fingerprint of encrypted data. Elevated kurtosis values may denote a distribution with pronounced tails or sharper peaks, characteristics potentially representative of encrypted data. In other words, elevated Byte Kurtosis in an unencrypted file may hint at a concentrated distribution of byte values. In contrast, for an encrypted file, Byte Kurtosis would typically approach zero, reflecting the uniform distribution of byte values.

\begin{align}
 {K} & = \frac{N(N+1)}{(N-1)(N-2)(N-3)}\frac{1}{N}\sum_{i=1}^{N}\left(\frac{x_i-\mu}{\sigma}\right)^4 & - \frac{3(N-1)^2}{(N-2)(N-3)}
\end{align}

\item \textbf{Byte Skewness:} This measure of distributional asymmetry can assist in differentiating between the recurring patterns found in unencrypted data and the randomness intrinsic to encrypted data. Encrypted data, due to its random characteristics, is anticipated to exhibit a skewness value approximating zero, while unencrypted data may manifest significant skewness if it contains repeated elements or patterns. 

\begin{align}
     S & = \frac{N(N-1)}{(N-2)}\frac{1}{N}\sum_{i=1}^{N}\left(\frac{x_i-\mu}{\sigma}\right)^3
\end{align}

In Variance, skewness, and kurtosis of byte content, $x_i$ is the byte value, $N$ is the total number of bytes, $\mu$ is the mean of the byte values, and $\sigma$ is the standard deviation of the byte values.

\item \textbf{Strings (Average String Length):} The diversity in string lengths within unencrypted files can provide hints regarding their nature. Encrypted files may manifest a uniform distribution of strings, whereas unencrypted files may exhibit varied string lengths attributable to the presence of natural language structures or formatting.

\begin{align}
    \text{avg\_string\_length} & = \frac{1}{N}\sum_{i=1}^{N}\left|s_i\right|
\end{align}
In Average string length, $s_i$ is a string in the file and $N$ is the total number of strings.

\item \textbf{File Content Analysis:} This feature encompasses a detailed examination of the file content, such as the frequency of specific patterns or the presence of distinct structures (e.g., headers, footers, metadata), providing additional clues that can assist in differentiating between encrypted and unencrypted files.

\begin{align}
    \text{avg\_word\_length} & = \frac{1}{N}\sum_{i=1}^{N}\left|w_i\right|
\end{align}

In Average word length, $w_i$ is a word in the file and $N$ is the total number of words.

\item \textbf{Entropy Variance:} By measuring the fluctuation in randomness across different sections of a file, this feature can aid in identifying encrypted files. Encrypted files may exhibit a comparatively consistent entropy level, whereas unencrypted files may show greater entropy variability. Note that the entropy is a measure of the disparity of the density function $f ( x )$ from the uniform distribution. On the other hand, the variance measures an average of distances of outcomes of the probability distribution $f ( x )$ from the mean \cite{toomaj2020generalized}.

\begin{align}
    \sigma_H^2 & = \frac{1}{N}\sum_{i=1}^{N}(H(X_i) - \mu_H)^2
\end{align}

 In Entropy variance, $H(X_i)$ is the entropy of the $i$-th 512-byte block, $N$ is the total number of 512-byte blocks, and $\mu_H$ is the mean entropy of the blocks.

\item \textbf{Percentiles utilization:} We assume that percentiles serve to appraise the dispersion of byte values within a file. Within a data cohort, the Nth percentile is characterized as the value beneath which N percent of the data resides. Hence, the 25th, 50th, and 75th percentiles epitomize the values beneath which 25\%, 50\%, and 75\% of the data reside, respectively. Plaintext files predominantly possess a structured framework, engendering the recurrence of particular byte values. For instance, within a text file, ASCII values corresponding to common alphabets and spaces occur with high frequency.
In the event of plotting the distribution of byte values, one would observe peaks at certain values, thereby inducing a skew in the distribution. As a consequence, the 25th, 50th, and 75th percentiles in a plaintext file may not adhere to a uniform distribution owing to these peaks and troughs. 

Encryption algorithms are conceived to metamorphose plaintext into a semblance that radiates randomness to any entity devoid of the decryption key. Optimally, this transformation engenders a uniform distribution of byte values spanning the feasible range (0 to 255 for 8-bit bytes). In such a distribution, the likelihood of each byte value is equitably distributed, and thus, the distribution's percentiles would settle at regular intervals across this range.
To elucidate:
\begin{itemize}
    \item The 25th percentile would approximate the value 64 (since 25\% of 256 is 64).
    \item The 50th percentile (also recognized as the median) would approximate the value 128 (since 50\% of 256 is 128).
    \item The 75th percentile would approximate the value 192 (since 75\% of 256 is 192).
\end{itemize}

This percentile analysis conjecturally can offer a proper feature to distinguish between normal and encrypted files predicated on the distribution of their byte values at the file level system.

\end{enumerate}

\subsubsection{Feature extraction computational resources}

The process of extracting statistical features from individual files, as detailed in the data, involves varying degrees of computational resource utilization, which is dependent on the file type and whether it is in its normal or encrypted state. For normal files, the resource demand is largely influenced by the file's size and type. For example, feature extraction from a large Excel file is considerably resource-intensive, requiring up to 441.827 seconds and utilizing 0.0352 MB of memory. Conversely, simpler files such as JPEG images and mp3 audio files demand significantly less computational power, with JPEG extraction taking a mere 0.0090 seconds and mp3 files requiring 0.927 seconds, both utilizing a minimal memory footprint of 0.0039 MB.

When files are encrypted by ransomware, the computational requirements for feature extraction exhibit slight variations. Some file types show a decrease in processing time, albeit with an increment in memory usage. For instance, the memory requirement for an encrypted Excel file slightly increases to 0.0391 MB. Table \ref{tab:feature_extraction} illustrates the differences in time and memory usage for extracting information from various file types, both in their normal state and when encrypted by ransomware.

\begin{table}[ht]
\centering
\caption{Computational resources required for feature extraction from normal and encrypted files by ransomware.}
\label{tab:feature_extraction}
\rowcolors{2}{Gray}{white} 
\resizebox{8cm}{!}{%
\begin{tabular}{lccc}
\hline
\rowcolor{white} 
\textbf{File Type} & \textbf{Size} & \textbf{Time (Second)} & \textbf{Memory Usage (MB)} \\
\hline
PDF Normal & 82.1 MB & 11.526 & 0.0039 \\
mp4 Normal & 182 MB & 26.845 & 0.0039 \\
JPEG Normal & 23.7 KB & 0.0090 & 0.0039 \\
Excel Normal & 3.58 GB & 441.827 & 0.0352 \\
Doc Normal & 89.8 MB & 12.772 & 0.0039 \\
mp3 Normal & 6.59 MB & 0.927 & 0.0039 \\
PDF Encrypt & 82.1 MB & 11.215 & 0.0156 \\
mp4 Encrypt & 182 MB & 26.069 & 0.0155 \\
JPEG Encrypt & 24.0 KB & 0.0040 & 0.0039 \\
Excel Encrypt & 3.58 GB & 421.884 & 0.0391 \\
Doc Encrypted & 89.8 MB & 12.760 & 0.0195 \\
mp3 Encrypted & 6.59 MB & 0.924 & 0.0039 \\
\hline
\end{tabular}}
\end{table}

\subsection{Benchmarking supervised learning algorithms for classifications: A comparative analysis}

In the empirical analysis, we employed a series of batch classifiers, as delineated in the preceding section, to assess the efficacy of both traditional shallow and deep learning algorithms in discerning encrypted from non-encrypted files across multiple file formats. Subsequently, the feature vectors were extracted for both the normal and encrypted files using our supervised machine learning tool, leading to a comprehensive dataset consisting of 65.1 GB (70,008,138,878 bytes) files.

Quantitative metrics are showcased in Table \ref{tab:classifier_performance11}, furnishing a comparative performance analysis of the various machine learning classifiers under review. These metrics encompass Accuracy, Precision, Recall, and F1-score, providing a thorough understanding of each model's performance capabilities. Notably, the training of these classifiers was rigorously based on the feature variables outlined in section \ref{sec:feature}. Our findings suggest that certain machine learning classifiers, particularly Random Forest, Decision Threes, AdaBoost, and Gradient Boosting, offer exceptional capabilities in the classification of files based on their encryption status, thereby providing valuable insights into effective strategies for the detection of ransomware-encrypted files. We partitioned the dataset, allocating 70\% for training and the remaining 30\% for evaluation. However, our assumption is that online learning, when incorporated with Decision Trees, can be suitable in such a file system (i.e., data stream) environment for learning and predicting encrypted and unencrypted buffers.

\begin{table}[h!] 
\centering

\caption{The table presents a comparative assessment of multiple machine learning classifiers, including Logistic Regression, Support Vector Machines (SVM), Decision Trees, Random Forests, k-Nearest Neighbors, AdaBoost, Gradient Boosting, Multilayer Perceptron, and Naive Bayes, in classifying files as encrypted or normal. Decision Trees emerged as the most accurate classifier, with an accuracy of 98.08\% and an F1-score of 98.20\%. Random Forests and Gradient Boosting also showcased robust performance, closely following Decision Trees with comparable precision and recall. These results underscore the effectiveness of Decision Trees, Random Forests, and Gradient Boosting in maintaining a harmonious balance between precision and recall, affirming their position as leading classifiers in this specific classification task.}\label{tab:classifier_performance11}

\resizebox{8cm}{!}{%
\begin{tabular}{
  l
  S[table-format=1.4]
  S[table-format=1.4]
  S[table-format=1.4]
  S[table-format=1.4]
}
\toprule
\rowcolor{mygray}
\textbf{Classifier} & {\textbf{Accuracy}} & {\textbf{Precision}} & {\textbf{Recall}} & {\textbf{F1}} \\
\midrule
Logistic Regression & 0.9155 & 0.8748 & 0.9878 & 0.9246 \\
\rowcolor{mygray}
SVM & 0.9502 & 0.9324 & 0.9793 & 0.9535 \\
Decision Trees & \textbf{0.9808} & \textbf{0.9496} & \textbf{0.9880} & \textbf{0.9820} \\
\rowcolor{mygray}
Random Forests & \textbf{ 0.9806} & \textbf{0.9703} & \textbf{0.9947} & \textbf{0.9786} \\
k-Nearest Neighbors & 0.9353 & 0.9226 & 0.9591 & 0.9385 \\
\rowcolor{mygray}
AdaBoost & 0.9719 & 0.9607 & 0.9879 & 0.9732 \\
Gradient Boosting & \textbf{0.9800} & \textbf{0.9686} & \textbf{0.9941} & \textbf{0.9808} \\
\rowcolor{mygray}
Multilayer Perceptron & 0.7897 & 0.9160 & 0.9801 &  0.8395 \\
Naive Bayes & 0.8918 & 0.8346 & 0.9967 & 0.9055\\
\bottomrule
\end{tabular}
}
\end{table}

The necessity for online incremental learning approaches to improve performance, even with such high accuracy rates presented in Table \ref{tab:classifier_performance11}, can be justified on several fronts:
\begin{enumerate}
    \item \textbf{File system Dynamic Environments:}In practical scenarios, the characteristics of files, user behavior, and system environments are not static. An online learning approach can adjust to changes in the environment that might affect the model's performance, such as new file types or legitimate encryption practices by users, reducing false positives and improving detection accuracy over time.
    \item \textbf{Model Drift:} Over time, the performance of models trained in an offline setting might degrade due to model drift. This phenomenon occurs when the underlying data distribution changes, making previously learned patterns less applicable. Online learning continuously updates the model, mitigating the effects of model drift and maintaining high accuracy levels.
    \item \textbf{Resource Efficiency:} Online learning models can be more resource-efficient in certain contexts. They require less memory and computational power to update with new data, compared to retraining an entire offline model with a growing dataset. This efficiency is particularly valuable in environments with limited resources.
\end{enumerate}

\subsection{Exploring the efficacy of incremental learning in predicting encrypted and unencrypted buffer at file system} \label{sec: Online Learning}
In traditional machine learning paradigms, particularly within supervised learning contexts, the modus operandi is predominantly batch-oriented. In this framework, an accumulated dataset serves as the initial substrate for the algorithmic training process. This necessitates the availability of the entire training dataset in advance of the model's learning phase, often relegating the training process to offline execution due to the substantial computational and temporal overheads involved \cite{HOI2021249}. Despite their ubiquity, batch learning techniques are fraught with inherent limitations, including (a) suboptimal efficiency concerning both time and computational resources, and (b) a lack of scalability, especially in large-scale applications where the introduction of new data typically mandates a comprehensive retraining of the existing model.

Incremental online learning represents a significant advancement in machine learning, offering unparalleled adaptability, efficiency, and real-time processing capabilities, especially in dynamic environments like file systems. This document addresses the concerns regarding its advantages over traditional ML schemes.

\begin{itemize}
    \item \textbf{Adaptability to New Data}: Incremental online learning algorithms are designed to adapt to new data in real-time or near-real-time, without the need for retraining the model from scratch \cite{8603156}. This is particularly advantageous in dynamic environments like file systems, where new file types may be introduced over time, and the system must quickly adapt to recognize and process these new structures. Traditional ML schemes, in contrast, often require batch processing and retraining on the entire dataset, including both old and new data, which is computationally expensive and time-consuming.

    \item \textbf{Efficiency in Handling Large Datasets:} File systems typically contain a vast amount of data. Incremental online learning algorithms can process data in smaller chunks, making them more memory-efficient than traditional approaches that may require the entire dataset to be loaded into memory. This efficiency is crucial for learning file types' structures and predicting data encryption status, where models must scale to accommodate large datasets without significant increases in computational resources.

    \item \textbf{Real-time Prediction and Detection:} The ability to update models incrementally allows for real-time prediction and detection of encrypted versus unencrypted data. This is essential for cybersecurity applications where timely detection of encrypted files (potentially indicative of ransomware activity) is critical. Traditional ML models, due to their batch-learning nature, cannot easily provide real-time insights or adapt quickly to newly emerging encryption patterns.

    \item \textbf{Experimental Comparison and Justification:} To empirically demonstrate the advantages of incremental online learning, we propose a comprehensive experimental comparison involving several datasets representative of real-world file system structures, including a mix of encrypted and unencrypted data.

    \item \textbf{Application-Specific Benefits:} For file system monitoring and encryption detection, the incremental learning approach allows for continuous learning from incremental changes in file structures or content, enhancing the model's ability to detect subtle shifts towards encryption patterns. This contrasts with traditional ML models that might miss or delay detecting these patterns due to the static nature of their training process.
    
\end{itemize}

In contrast, online learning provides a more adaptive machine learning paradigm well-suited for data streams that are sequentially ordered. The underlying objective is the iterative refinement and dynamic updating of the model to maximize predictive performance on future, as-yet-unseen data. This adaptability effectively ameliorates the limitations of batch learning by enabling incremental model updates as new instances of data become available. Consequently, online learning algorithms demonstrate enhanced efficiency and scalability, attributes particularly beneficial in large-scale, real-world data analytics scenarios where data is not only voluminous but also arrives at a high velocity \cite{HOI2021249} \cite{8954008}.

\subsection{Computational cost comparison between the traditional and online learning models}
The comparison between the traditional and the online learning models, based on evaluations conducted using our full dataset, reveals significant differences in computational efficiency and resource utilization. The Traditional Model, while delivering high accuracy, requires a considerably longer CPU time of approximately 3.76 seconds and consumes around 7.47 MB of memory. This indicates a substantial computational effort and resource allocation to achieve its predictive performance, highlighting the model's reliance on the extensive dataset for training and evaluation. On the other hand, the online learning model stands out for its remarkable efficiency, needing only about 0.27 seconds of CPU time and a minimal memory usage of 0.016 MB. Despite the differences in accuracy metrics, the online learning model's performance showcases its potential for applications requiring rapid responses and minimal resource consumption. The utilization of the full dataset for obtaining these results underscores the efficiency of the online learning model in handling large volumes of data swiftly and with significantly lower resource requirements, making it particularly suitable for real-time or resource-constrained environments. Table \ref{tab:PerformanceMetrics} compares the performance of a traditional machine learning model, specifically, a Random Forest Classifier, and an online learning model, specifically, SGDClassifier in terms of CPU time and memory usage.

\begin{table}[ht]
\centering
\caption{Computational costs: The traditional model is more resource-intensive, while the online learning model is highly efficient, demonstrating lower CPU time and minimal memory usage, highlighting its suitability for real-time applications with strict resource constraints.}\label{tab:PerformanceMetrics}
\resizebox{8cm}{!}{%
\begin{tabular}{lcc}
\toprule
Model & CPU Time (s) & Memory Usage (MB) \\
\midrule
Traditional & 3.755 & 7.473 \\
Online Learning & 0.266 & 0.016 \\
\bottomrule
\end{tabular}}
\end{table}

\subsubsection{Online learning benchmarks}
To rigorously assess the most effective machine learning classifiers for distinguishing between encrypted and unencrypted files, we conducted a comprehensive empirical study using a large and diverse dataset. The dataset featured 11,928 files, totaling 32.6GB in size, encrypted through various techniques from 75 unique ransomware families. Noteworthy among these were sophisticated encryption strategies like partial and intermittent encryption, commonly employed by ransomware variants such as Black Basta, BlackMatter, Grand Crab, lab-developed ransomware AES-Base64, Paradise, and LockBit 3.0.

\subsection{First-order online learning }

In the context of online learning, First-order algorithms are particularly useful for large-scale learning tasks and data streams due to their incremental model updating using only gradient information (i.e., first-order derivative). In our investigation, we leveraged conventional machine learning classifiers that fall under the category of first-order online learning algorithms. These include a  \textit{Stochastic Gradient Descent (SGD) (logistic regression as the linear classifier)}, \textit{Passive-Aggressive techniques}, and the \textit{Perceptron model}. These algorithms were strategically employed for predictive analytics on a file system-based dataset, enabling us to rigorously explore the inherent differences between categories of encrypted and unencrypted data within a complex data corpus for large-scale learning tasks and data streams.

\subsubsection{Mathematical formulation in an online learning context}

Let  $\mathcal{D} = \{(x_1, y_1), (x_2, y_2), \ldots, (x_n, y_n)\}$ represent a heterogeneous dataset sourced from a file system. Here, each \(x_i\) denotes a data buffer, and \(y_i \in \{0, 1\}\) signifies whether the buffer is encrypted (1) or unencrypted (0). We applied machine learning classifiers, particularly focusing on algorithms such as Stochastic Gradient Descent (SGD), Passive-Aggressive algorithms, and the Perceptron model. The algorithms aim to minimize the objective function \(J(\theta)\) concerning the model parameters \(\theta\).


For the SGD algorithm, the update rule at time \(t\) is as follows:
\[
\theta_{t+1} = \theta_t - \alpha_t \nabla J(\theta_t)
\]


In the case of the Passive-Aggressive algorithm, the update rule can be defined as:
\[
\theta_{t+1} = \theta_t + \tau_t (y_t - \text{sgn}(\theta_t^T x_t)) x_t
\]
Here, \(\tau_t\) is selected to minimize \(J(\theta)\), subject to specific constraints.


For the Perceptron algorithm, the update occurs conditionally upon an error:
\[
\theta_{t+1} = \theta_t + \alpha_t (y_t - \text{sgn}(\theta_t^T x_t)) x_t
\]
This update is executed only when \(y_t \neq \text{sgn}(\theta_t^T x_t)\).

utilizing these algorithms, we continually adjust \(\theta\) with each incoming sample \((x_t, y_t)\).

The dynamic adaptability and real-time learning capabilities of these algorithms render them highly effective for the real-time detection of encrypted versus unencrypted data buffers at the file system level. The methodology thus provides a detailed and nuanced understanding of the intrinsic variances between encrypted and unencrypted data, making it possible to detect encrypted buffers in real time. This paves the way for detecting between encrypted and unencrypted data buffers.

\subsection{Hoeffding Tree algorithm and key advantages}

The Hoeffding Tree Algorithm, also known as VFDT (Very Fast Decision Tree Learner), serves as a cutting-edge approach for the incremental induction of decision trees from unbounded data streams. This algorithm is designed to analyze each incoming data instance exactly once, thereby eliminating the need to store past instances once they have been incorporated into the evolving decision tree. Within this computational framework, the only entity that necessitates in-memory storage is the decision tree itself, which accumulates relevant metadata within its terminal nodes to not only allow for future growth but also to enable real-time predictive inference. The mathematical robustness of the Hoeffding Tree Algorithm has been formally corroborated by Domingos and Hulten, who have demonstrated that the algorithm produces decision trees closely approximating those generated through conventional batch learning methods \cite{Hulten}. This finding lends empirical weight to the argument that the Hoeffding Tree is highly efficacious in producing high-quality decision trees, thereby aligning with the prevailing consensus that batch-learned decision trees are among the most effective model architectures in the field of machine learning. Here are some key advantages, especially in scenarios involving streaming data, such as a file system space:

\begin{itemize}
    \item \textbf{Adaptability to Heterogeneous Data:} In environments where data characteristics can change over time due to various factors such as user behavior, system updates, or evolving threat landscapes, the ability of online learning models like the Hoeffding tree to adapt and learn incrementally from each new batch of data becomes invaluable. This adaptability ensures that the model remains effective even as the nature of the data it analyzes changes, providing a level of resilience that static offline models may lack.
    
    \item \textbf{Resource Efficiency:} One of the significant advantages of the Hoeffding tree, highlighted in the performance metrics, is its efficiency in terms of CPU and memory usage. This efficiency is critical in real-time file system environments where resources are limited, and the overhead of running complex detection models can be prohibitive. The ability of the Hoeffding tree to deliver effective performance while minimizing resource consumption makes it particularly suitable for deployment in such contexts, where maintaining system performance and responsiveness is paramount.

    \item \textbf{Real-Time Detection Capabilities:} The Hoeffding tree's online incremental learning approach enables it to update its knowledge base in real-time as new data arrives. This capability is crucial for the timely detection of ransomware attacks, where early detection and response can significantly mitigate the impact. In contrast, offline models, despite their high accuracy, require periodic retraining and deployment, which can introduce delays in adapting to new threats.

    \item \textbf{Continuous Improvement and Learning:} Although the performance of the Hoeffding tree may appear to stabilize after a certain point, this does not preclude the potential for future improvements as more diverse data is encountered. The model's design to shuffle and learn from the data in a way that reflects the heterogeneous nature of file systems ensures that it remains sensitive to subtle changes in data patterns that could indicate new or evolving threats.

    \item \textbf{Scalability:} The scalability of online learning models like the Hoeffding tree, given their incremental nature, is another factor that contributes to their standout performance. As the volume of data in file systems grows, the model can continue to learn and adapt without the need for complete retraining, making it well-suited to environments with expanding data volumes.
    
\end{itemize}

\subsection{Online learning evaluations and results}

We have developed a tool designed to evaluate and compare online learning algorithms, specifically addressing the challenges presented by large-scale or real-time data within environments like enterprise-level corporate offices or multi-tenant cloud storage. This tool employs an incremental learning paradigm, leveraging a sliding window mechanism to manage the data files. This operational approach enables the incremental training of classifiers such as SGD (Stochastic Gradient Descent), Perceptron, Passive Aggressive, and Hoeffding Tree, making it exceptionally suitable for situations where the data's size prohibits in-memory storage or when data is generated in real-time, such as at the file system level. By reading, processing, and using a subset of the data for model training during each computational iteration, the tool overcomes limitations related to computational resources. This strategy allows for the dynamic ingestion of data, ensuring that the system remains efficient and effective even as data volumes grow or as new data streams in continuously. 

To assess the effectiveness of online learning models, we encrypted a dataset with 75 ransomware families using a lab-developed AES-Base64 encryption-encoding model. The results, presented in Table \ref{tab:ComparisonTable1}, showcase the performance of these online learning models when confronted with well-known ransomware, alongside the encryption-encoding model. Here brief descriptions of each online learning evaluation conducted against the encrypted dataset are provided.

\paragraph{\textbf{GrandCrab Ransomware evaluation:}}
GandCrab, a notorious ransomware, has significantly impacted global cybersecurity by encrypting user files and demanding ransom payments for decryption keys. It distinguished itself through its ability to evade detection with frequent updates and a variety of distribution methods. In our study, we encrypted data using the latest GandCrab variant to test the efficacy of online learning models. The results illustrated that these models frequently achieve over 90\% accuracy in distinguishing between normal and encrypted files. Particularly noteworthy is the performance in Batch 4, where the Hoeffding Tree classifier achieved perfect classification (accuracy = 100\%). This suggests that the Hoeffding Tree is exceptionally effective at capturing the underlying data distribution, making it a potent tool for identifying and mitigating threats posed by sophisticated ransomware like GandCrab. Result illustrated in Table \ref{tab:ComparisonTable1} row A.

\paragraph{\textbf{AES-Base64 encryption-encoding evaluation (lab-developed):}} 
One malicious method to reduce entropy involves using AES-Base64 for encryption and encoding. We developed a tool for encrypting and encoding data into Base64 and evaluated its effectiveness with online learning algorithms. Our empirical experiment, detailed in the second row (B) of Table \ref{tab:ComparisonTable1}, outlines the performance of online  learning models. The Hoeffding Tree classifier stands out for its proficiency, especially from batch 2 onwards, consistently scoring above 95\% and achieving a perfect score of 1.0 in batch 5. This superior performance suggests its enhanced adaptability to evolving data or superior capability in online learning. Conversely, the Passive Aggressive classifier peaks at a score of 77.5\% in batch 11 but shows variability across other batches, indicating potential sensitivity to data distribution. The SGD and Perceptron classifiers exhibit moderate and varying performances, with SGD reaching approximately 72\% in batch 12 and Perceptron peaking at 65\% in batches 3 and 4.

\paragraph{\textbf{Paradise Ransomware evaluation (Partial encryption):}} Paradise ransomware has become a notable threat in the cybersecurity domain, characterized by its advanced encryption methods and the severe impact on the data of its victims. Through reverse-engineering analysis, it has been found that Paradise ransomware conducts two primary operations on the files it targets: reading and encrypting the content. The encryption strategy of Paradise is meticulously tailored to the size of the files it infects. Specifically, files less than 5MB in size are entirely encrypted by the ransomware, rendering them completely inaccessible. For files that fall within the 5MB to 100MB range, Paradise employs partial encryption. This involves encrypting 5MB of the file's content, divided into two sections of 2.5MB each, with one section from the beginning and another from the end of the file. Files exceeding 100MB undergo a similar approach of partial encryption, where 25MB of the content is encrypted. However, this is done in a staggered fashion, with the encryption distributed across ten 2.5MB chunks, each positioned at intervals spanning every 10\% of the file's total size \cite{cocomazzi2022}. 

Our empirical evaluation highlights the Hoeffding Tree classifier's exceptional and consistent performance in distinguishing between encrypted and normal data, particularly evident from Batch 2 onwards. It consistently scores above 95\% accuracy, reaching a perfect score in Batch 9. This consistent high performance underlines the classifier's adaptability and superior online learning capabilities in the context of encryption detection, as detailed in the results presented in Table \ref{tab:ComparisonTable1} Row C).

\begin{table*}[ht!]
    \centering
    \caption{ Comparative Performance of Online Learning Classifiers against Adversarial Encryption Techniques. This table illustrates the incremental evaluation results of various classifiers, with data fed in rounds of 1000, in a simulated file system-like environment. Four encryption techniques were analyzed: Grand Crab Ransomware’s Strong Encryption, AES-Base64’s Suggested Entropy Reduction, Paradise’s Partial Encryption, and CryptoFile2 ransomware which uses encrypted by a strong encryption with RSA-2048.} A particular focus is placed on the standout performance of the Hoeffding classifier, which exhibited superiority across all tested contexts. \textbf{Row A} illustrates the results from data encrypted with GrandCrab Ransomware. Row B illustrates data first encrypted with the AES algorithm and then encoded using Base64 encoding. \textbf{Row C} illustrates the results from data encrypted by Paradise ransomware, which uses partial encryption. {\textbf{Row D} illustrates the results from encrypted data by CryptFile2 ransomware. The Hoeffding algorithm shows outperforming accuracy results compared to other online learning algorithms. \label{tab:ComparisonTable1}
    
    \begin{tabular}{|l|c|c|}
        \hline
         \# & Comparison of Classifier Performance Graph & Comparison of Classifier Performance (Accuracy)\\
        \hline
         \makecell[l]{\textbf{A}} & \adjustbox{valign=c}{\includegraphics[width=6.5cm]{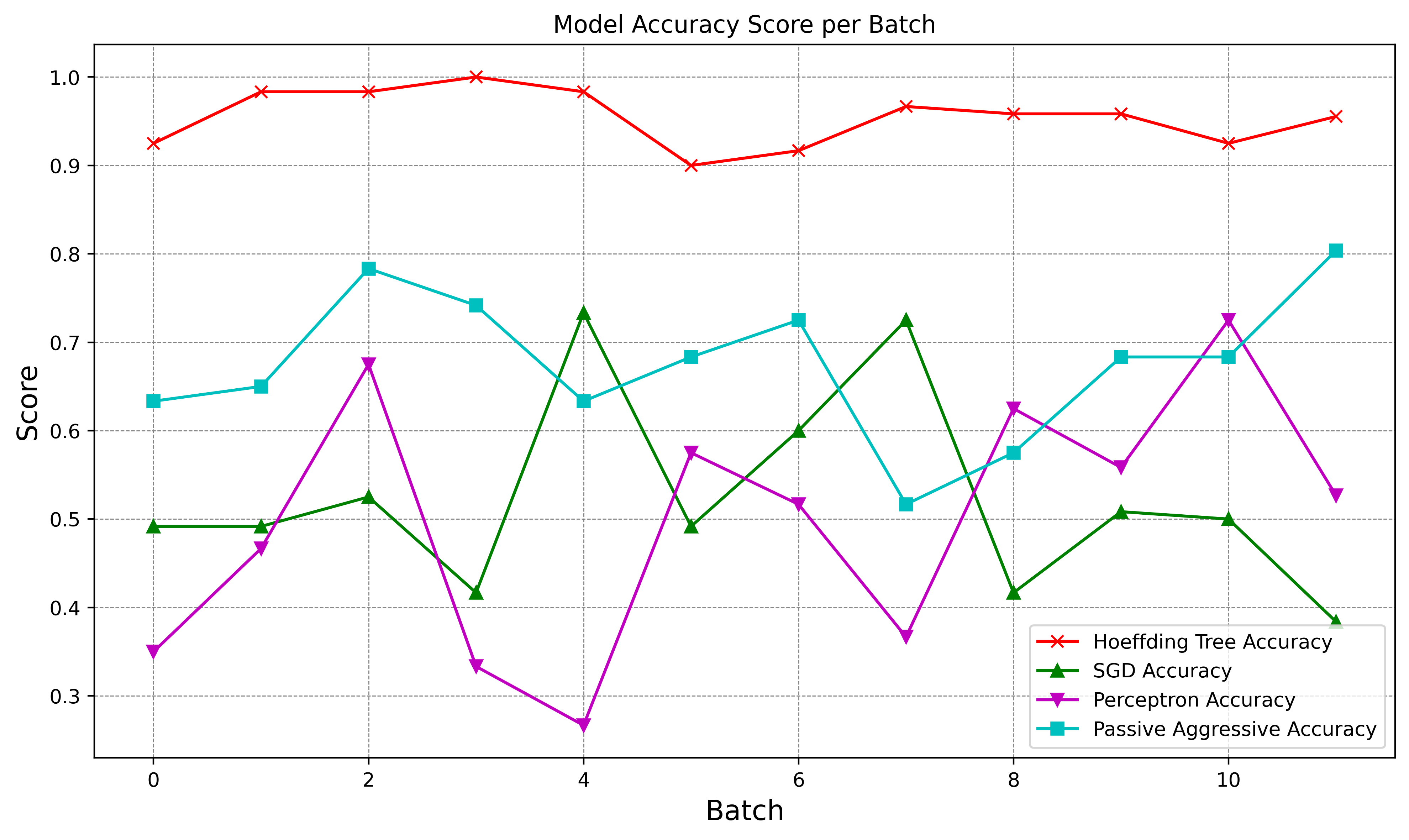}} &
        \adjustbox{valign=c}{ \resizebox{6cm}{!}{%
    \begin{tabular}{
        l
        S[table-format=1.4]
        S[table-format=1.4]
        S[table-format=1.4]
        S[table-format=1.4]
    }
        \toprule
        Batch & {SGD} & {Perceptron} & {Passive Aggressive} & {Hoeffding Tree} \\
        \midrule
        \rowcolor{mygray}
        1 & 0.4917 & 0.35   & 0.6333 & \textbf{0.9250} \\
        2 & 0.4917 & 0.4667 & 0.65   & \textbf{0.9833} \\
        \rowcolor{mygray}
        3 & 0.5250 & 0.6750 & 0.7833 & \textbf{0.9833} \\
        4 & 0.4167 & 0.3333 & 0.7417 & \textbf{1.0000} \\
        \rowcolor{mygray}
        5 & 0.7333 & 0.2667 & 0.6333 & \textbf{0.9833} \\
        6 & 0.4917 & 0.5750 & 0.6833 & \textbf{0.9000} \\
        \rowcolor{mygray}
        7 & 0.6000 & 0.5167 & 0.7250 & \textbf{0.9167} \\
        8 & 0.7250 & 0.3667 & 0.5167 & \textbf{0.9667} \\
        \rowcolor{mygray}
        9 & 0.4167 & 0.6250 & 0.5750 & \textbf{0.9583} \\
        10 & 0.5083 & 0.5583 & 0.6833 & \textbf{0.9583} \\
        \rowcolor{mygray}
        11 & 0.5000 & 0.7250 & 0.6833 & \textbf{0.9250} \\
        12 & 0.3839 & 0.5268 & 0.8036 & \textbf{0.9554} \\
        \bottomrule
    \end{tabular}
    }}  
        \\
        \hline 
        \makecell[l]{\textbf{B}}
         &
        \adjustbox{valign=c}{\includegraphics[width=6.5cm]{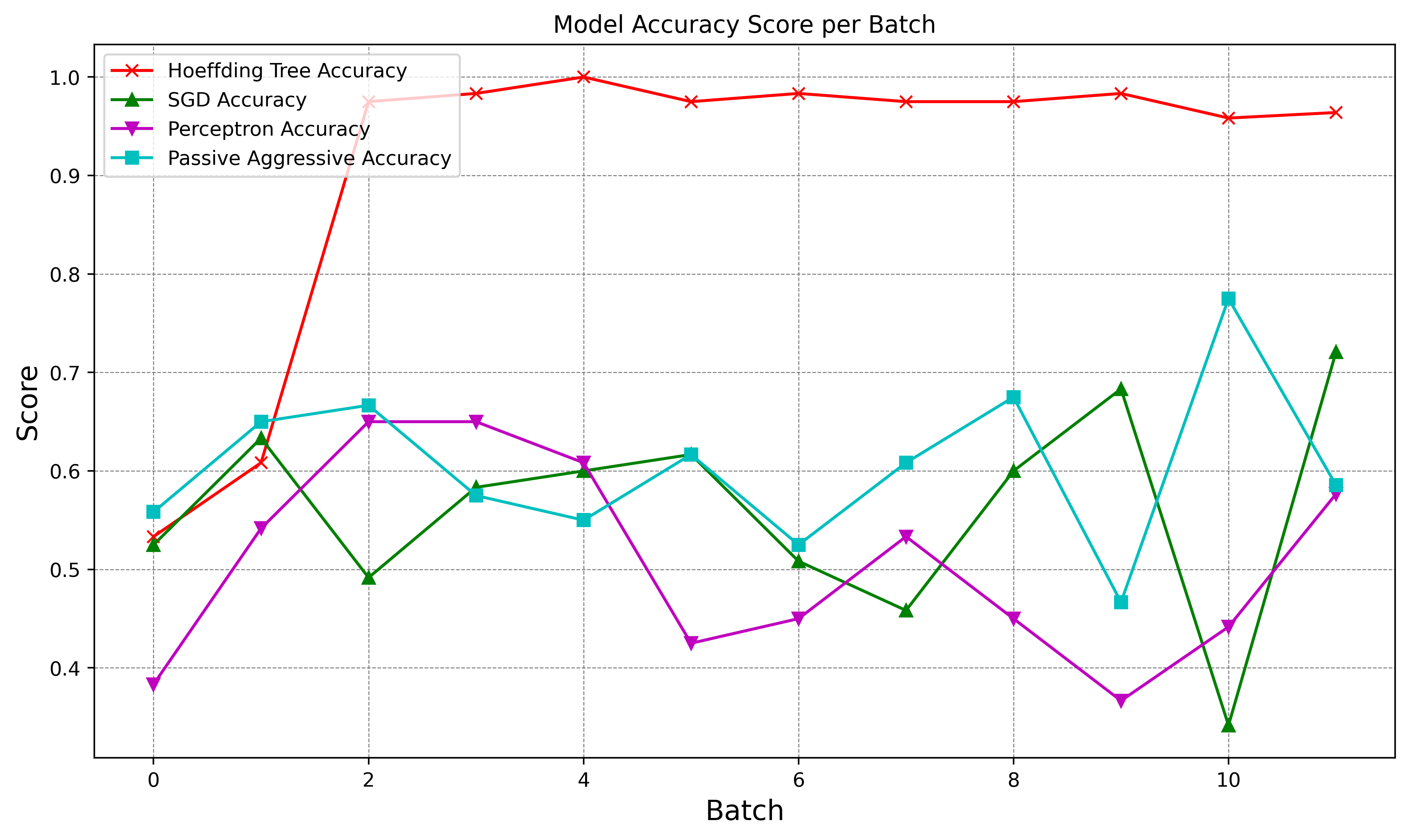}}  & 
        \adjustbox{valign=c}{\resizebox{6cm}{!}{%
    \begin{tabular}{
        l
        S[table-format=1.4]
        S[table-format=1.4]
        S[table-format=1.4]
        S[table-format=1.4]
    }
        \toprule
        Batch & {SGD} & {Perceptron} & {Passive Aggressive} & {Hoeffding Tree} \\
        \midrule
        \rowcolor{mygray}
        1 & 0.5250 & 0.3833 & 0.5583 & 0.5333 \\
        2 & 0.6333 & 0.5417 & 0.6500 & 0.6083 \\
        \rowcolor{mygray}
        3 & 0.4917 & 0.6500 & 0.6667 & \textbf{0.9750} \\
        4 & 0.5833 & 0.6500 & 0.5750 & \textbf{0.9833} \\
        \rowcolor{mygray}
        5 & 0.6000 & 0.6083 & 0.5500 & \textbf{1.0000} \\
        6 & 0.6167 & 0.4250 & 0.6167 & \textbf{0.9750} \\
        \rowcolor{mygray}
        7 & 0.5083 & 0.4500 & 0.5250 & \textbf{0.9833} \\
        8 & 0.4583 & 0.5333 & 0.6083 & \textbf{0.9750} \\
        \rowcolor{mygray}
        9 & 0.6000 & 0.4500 & 0.6750 & \textbf{0.9750} \\
        10 & 0.6833 & 0.3667 & 0.4667 & \textbf{0.9833} \\
        \rowcolor{mygray}
        11 & 0.3417 & 0.4417 & 0.7750 & \textbf{0.9583} \\
        12 & 0.7207 & 0.5766 & 0.5856 & \textbf{0.9640} \\
        \bottomrule
    \end{tabular}
    }}\\
    \hline
    \makecell[l]{\textbf{C}}
      &
    \adjustbox{valign=c}{\includegraphics[width=6.5cm]{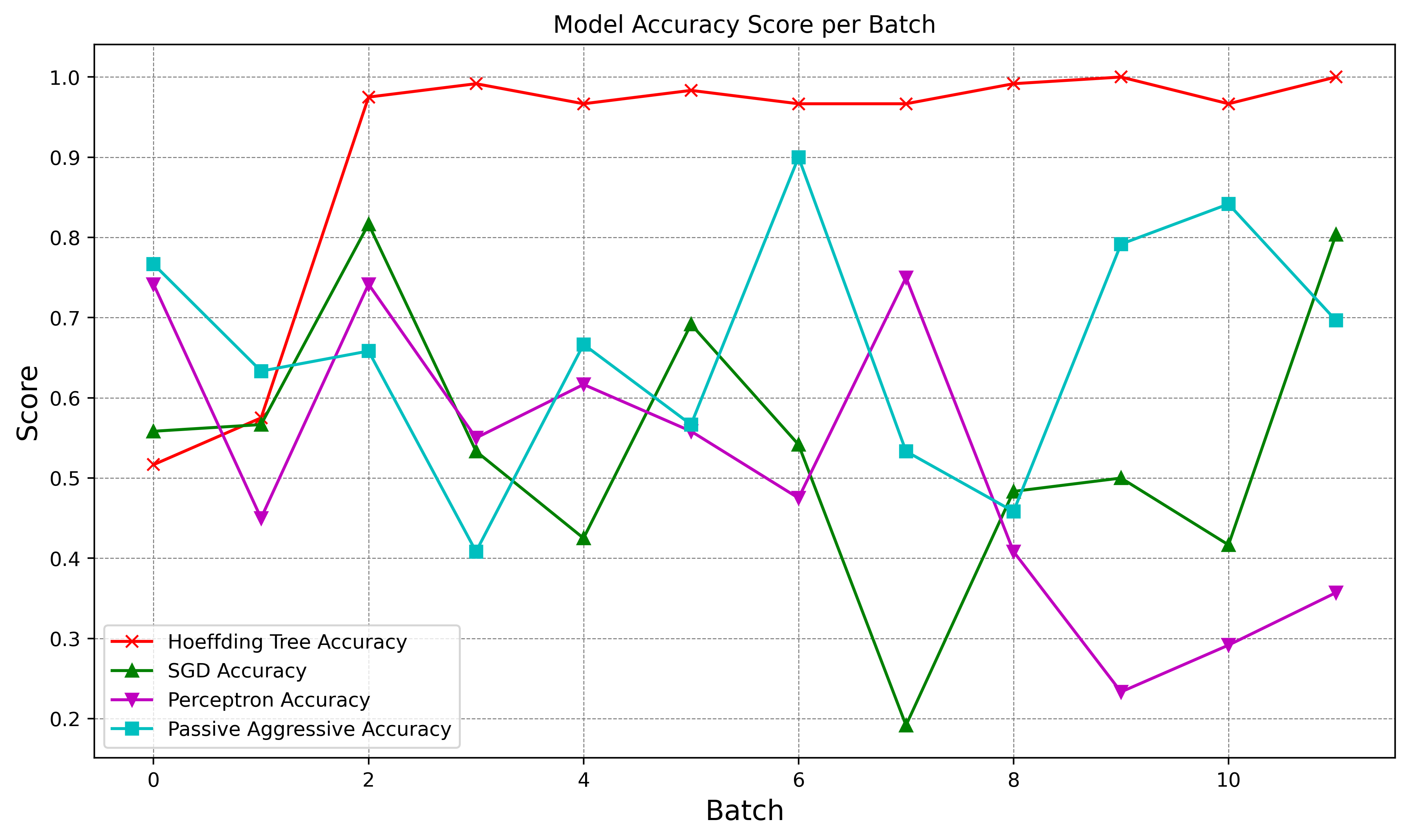}} &
    \adjustbox{valign=c}{\resizebox{6cm}{!}{%
    \begin{tabular}{
        l
        S[table-format=1.4]
        S[table-format=1.4]
        S[table-format=1.4]
        S[table-format=1.4]
    }
        \toprule
        Batch & {SGD} & {Perceptron} & {Passive Aggressive} & {Hoeffding Tree} \\
        \midrule
        \rowcolor{mygray}
        0 & 0.5583 & 0.7417 & 0.7667 & 0.5167 \\
        1 & 0.5667 & 0.4500 & 0.6333 & 0.5750 \\
        \rowcolor{mygray}
        2 & 0.8167 & 0.7417 & 0.6583 & \textbf{0.9750} \\
        3 & 0.5333 & 0.5500 & 0.4083 & \textbf{0.9917} \\
        \rowcolor{mygray}
        4 & 0.4250 & 0.6167 & 0.6667 & \textbf{0.9667} \\
        5 & 0.6917 & 0.5583 & 0.5667 & \textbf{0.9833} \\
        \rowcolor{mygray}
        6 & 0.5417 & 0.4750 & 0.9000 & \textbf{0.9667} \\
        7 & 0.1917 & 0.7500 & 0.5333 & \textbf{0.9667} \\
        \rowcolor{mygray}
        8 & 0.4833 & 0.4083 & 0.4583 & \textbf{0.9917} \\
        9 & 0.5000 & 0.2333 & 0.7917 & \textbf{1.0000} \\
        \rowcolor{mygray}
        10 & 0.4167 & 0.2917 & 0.8417 & \textbf{0.9667} \\
        11 & 0.8036 & 0.3571 & 0.6964 & \textbf{1.0000} \\
        \bottomrule
    \end{tabular}
    }}\\
    \hline 

    \makecell[l]{\textbf{D}}
      &
    \adjustbox{valign=c}{\includegraphics[width=6.5cm]{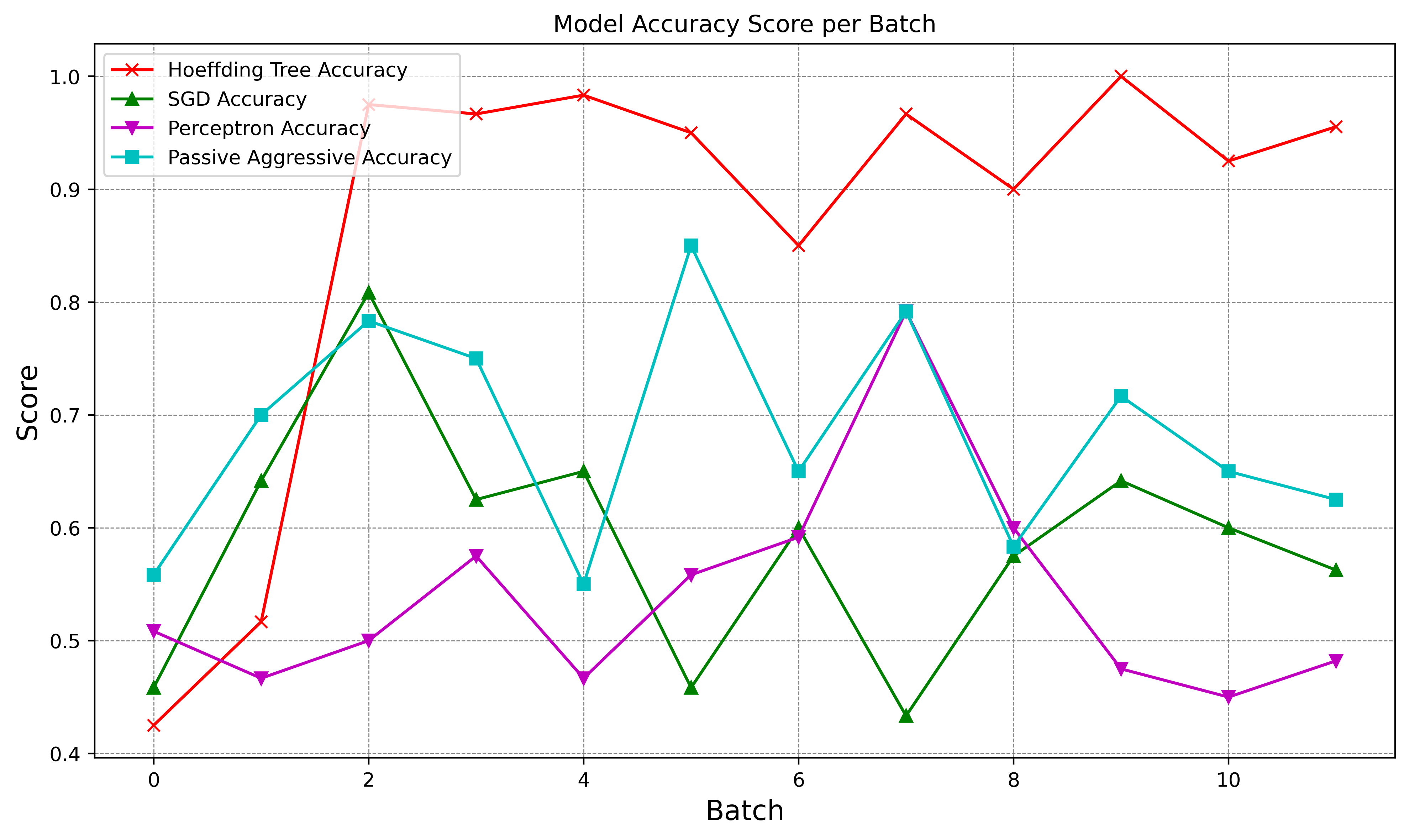}} &
    \adjustbox{valign=c}{\resizebox{6cm}{!}{%
    \begin{tabular}{
        l
        S[table-format=1.4]
        S[table-format=1.4]
        S[table-format=1.4]
        S[table-format=1.4]
    }
        \toprule
        Batch & {SGD} & {Perceptron} & {Passive Aggressive} & {Hoeffding Tree} \\
        \midrule
        \rowcolor{mygray}
        0 & 0.4583 & 0.5083 & 0.5583 & 0.4250 \\
        1 & 0.6416 & 0.4666 & 0.7000 & 0.5166 \\
        \rowcolor{mygray}
        2 & 0.8083 & 0.5000 & 0.7833 & \textbf{0.9750} \\
        3 & 0.6250 & 0.5750 & 0.7500 & \textbf{0.9666} \\
        \rowcolor{mygray}
        4 & 0.6500 & 0.4666 & 0.5500 & \textbf{0.9833} \\
        5 & 0.4583 & 0.5583 & 0.8500 & \textbf{0.9500} \\
        \rowcolor{mygray}
        6 & 0.6000 & 0.5916 & 0.6500 & \textbf{0.8500} \\
        7 & 0.4333 & 0.7916 & 0.7916 & \textbf{0.9666} \\
        \rowcolor{mygray}
        8 & 0.5750 & 0.6000 & 0.5833 & \textbf{0.9000} \\
        9 & 0.6416 & 0.4750 & 0.7166 & \textbf{1.0000} \\
        \rowcolor{mygray}
        10 & 0.6000 & 0.4500 & 0.6500 & \textbf{0.9250} \\
        11 & 0.5625 & 0.4821 & 0.6250 & \textbf{0.9553} \\
        \bottomrule
    \end{tabular}
    }}\\
    \hline     
    \end{tabular}}
\end{table*}

\paragraph{\textbf{Black Basta Ransomware evaluation (intermittent encryption):}} The advancement of intermittent encryption has changed the strategic adversarial landscape, favoring the attacker over the defender. By employing dynamic encryption and enhancing its stealth, adversaries can effectively bypass prevailing defense countermeasures. This is particularly true for systems that rely on I/O request packets (IRPs) to detect anomalous behavior. We encrypt the dataset using recent Black Basta variants. The Black Basta ransomware exhibits a unique encryption behavior that does not allow for operator-configurable encryption modes but rather implements intermittent encryption dictated by the size of the target file. 

\begin{figure}[h!]
\centering
\begin{tikzpicture}
    \node [above right,inner sep=0] (image) at (0,0) {\includegraphics[width=8cm]{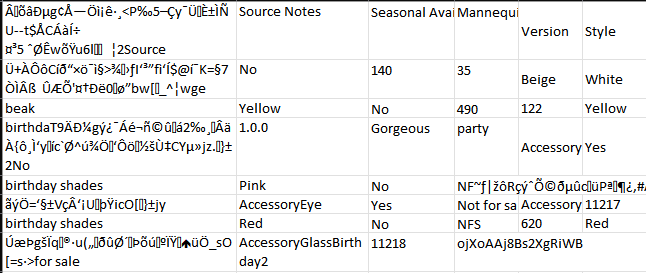}};
\begin{scope}[
    x={($0.1*(image.south east)$)},
    y={($0.1*(image.north west)$)}]
    \draw[very thick,green] (0,10) rectangle (3.6,8.9) node[above right,black,fill=green]{\tiny (1) First 64 encryption};
    \draw[very thick,red] (0,7.8) rectangle (3.6,8.7) node[below right,black,fill=green]{\tiny (3) Second 64 encryption after };
    \draw[latex-, very thick,green] (9.5,9) -- ++(-1,-3)  node[left,black,fill=green]{\scriptsize (2) Skipping 128 bytes before encrypting the next 64-byte};
\end{scope}
\end{tikzpicture}
\caption{Black Basta encrypts CSV files larger than 4 KB. The ransomware persistently encrypts 64-byte portions, maintaining a consistent interval of 128 bytes until reaching the end of the file. This is an example of how intermittent encryption impacts defender strategies that use I/O request packets. } \label{fig: Black_basta_64_192Bytes}
\end{figure}

Specifically, Black Basta undertakes a comprehensive encryption of the entire file content when the file size is less than 704 bytes. For files that are larger but remain under 4 KB in size, the ransomware encrypts in increments of 64 bytes, initiating from the beginning of the file and thereafter skipping 192 bytes before encrypting the next 64-byte block. The process alters when encrypting files exceeding 4 KB in size; here, Black Basta again encrypts 64-byte blocks starting from the file's inception but with a reduced skip interval of 128 bytes between each encrypted segment. Through a detailed analysis, it was discerned that in the case of files larger than 4 KB, the ransomware persistently encrypted 64-byte portions, maintaining a consistent interval of 128 bytes until reaching the file's end. This operational methodology elucidates a distinctive encryption mechanism employed by Black Basta, with the encryption pattern being solely influenced by the size of the file under encryption. Figure \ref{fig: Black_basta_64_192Bytes} illustrates how Black Basta encrypted csv file. 

\begin{table*}[h!]
    \centering
    \caption{Comparative Analysis of Online Classifiers Against Intermittent Encryption Adversarial Threats. This table delineates the performance distinctions among various online classifiers when contending with data encrypted using intermittent encryption. Initial observations reveal a suboptimal performance by the Hoeffding classifier in contrast to other encryption strategies. However, upon deeper examination, it emerges that the Random Forest classifier notably surpasses its counterparts, exhibiting a superior predictive accuracy concerning intermittent encryption. \textbf{Row A} illustrates the use of intermittent encryption, which results in underperforming the Hoeffding Tree. \textbf{Row B} illustrates benchmarking classifiers with differential entropy as an additional feature, resulting in the random forest classifier outperforming others. \textbf{Row C} illustrates a comparison between the Random Forest and Hoeffding Tree classifiers.}
    \label{tab:Comparetive Table 2}
    \begin{tabular}{|l|l|l|}
        \hline
         \textbf{\#} & Comparison of Classifier Performance Graph & Comparison of Classifier Performance  \\
        \hline
        \makecell[l]{\textbf{A}} & \adjustbox{valign=c}{\includegraphics[width=6.5cm]{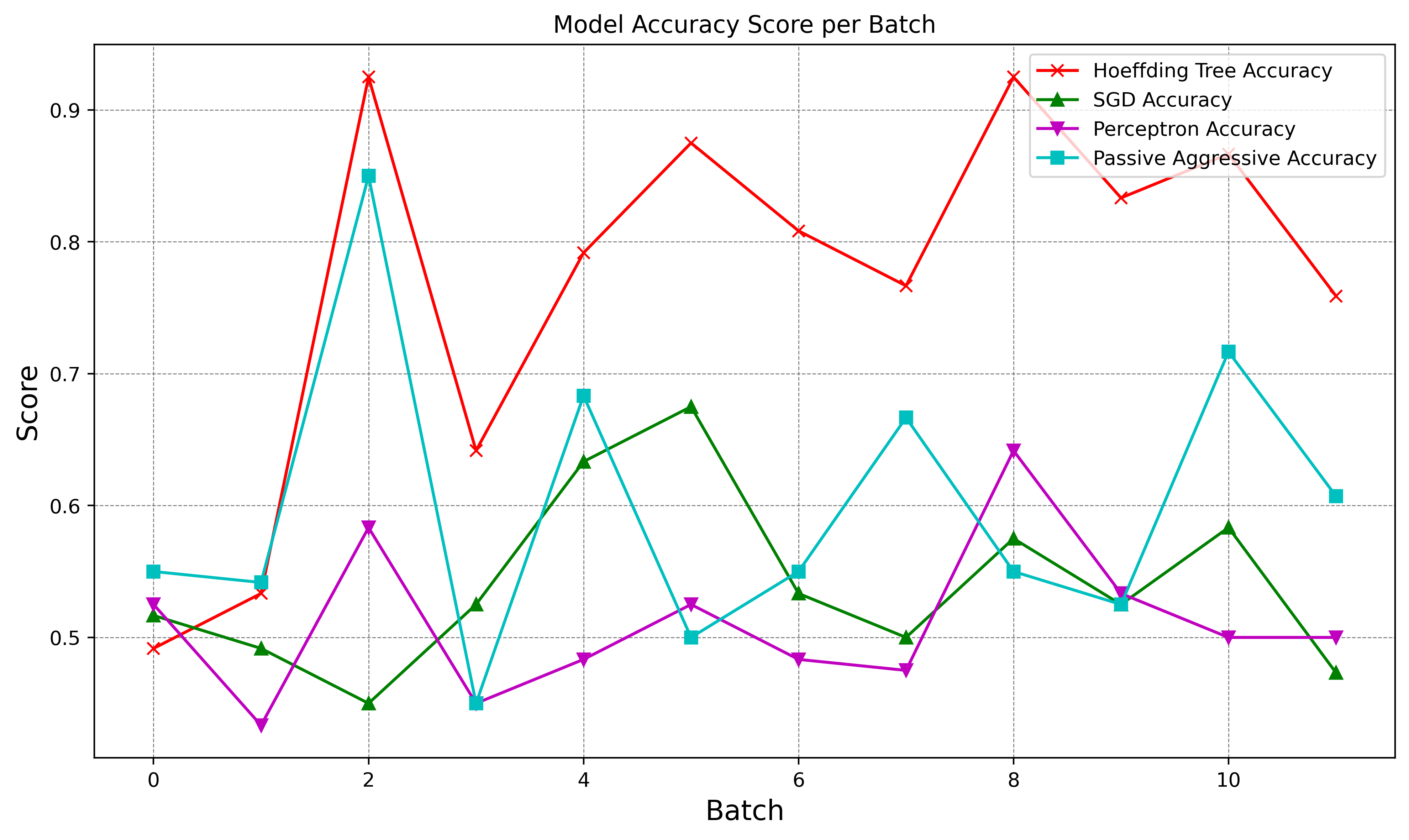}} &
        \adjustbox{valign=l}{\resizebox{6.6cm}{!}{%
    \begin{tabular}{
        l
        S[table-format=1.4]
        S[table-format=1.4]
        S[table-format=1.4]
        S[table-format=1.4]
    }
        \toprule
        Batch & {SGD } & {Perceptron } & {Passive Aggressive } & {Hoeffding Tree} \\
        \midrule
        \rowcolor{mygray}
        0 & 0.5167 & 0.5250 & 0.5500 & 0.4917 \\
        1 & 0.4917 & 0.4333 & 0.5417 & 0.5333 \\
        \rowcolor{mygray}
        2 & 0.4500 & 0.5833 & 0.8500 & \textbf{0.9250} \\
        3 & 0.5250 & 0.4500 & 0.4500 & 0.6417 \\
        \rowcolor{mygray}
        4 & 0.6333 & 0.4833 & 0.6833 & 0.7917 \\
        5 & 0.6750 & 0.5250 & 0.5000 & \textbf{0.8750} \\
        \rowcolor{mygray}
        6 & 0.5333 & 0.4833 & 0.5500 & 0.8083 \\
        7 & 0.5000 & 0.4750 & 0.6667 & 0.7667 \\
        \rowcolor{mygray}
        8 & 0.5750 & 0.6417 & 0.5500 & \textbf{0.9250} \\
        9 & 0.5250 & 0.5333 & 0.5250 & 0.8333 \\
        \rowcolor{mygray}
        10 & 0.5833 & 0.5000 & 0.7167 & \textbf{0.8667} \\
        11 & 0.4732 & 0.5000 & 0.6071 & 0.7589 \\
        \bottomrule
    \end{tabular}
    }}  
        \\
        \hline 
        \makecell[l]{\textbf{B}} & 
        \adjustbox{valign=c}{\includegraphics[width=6.5cm]{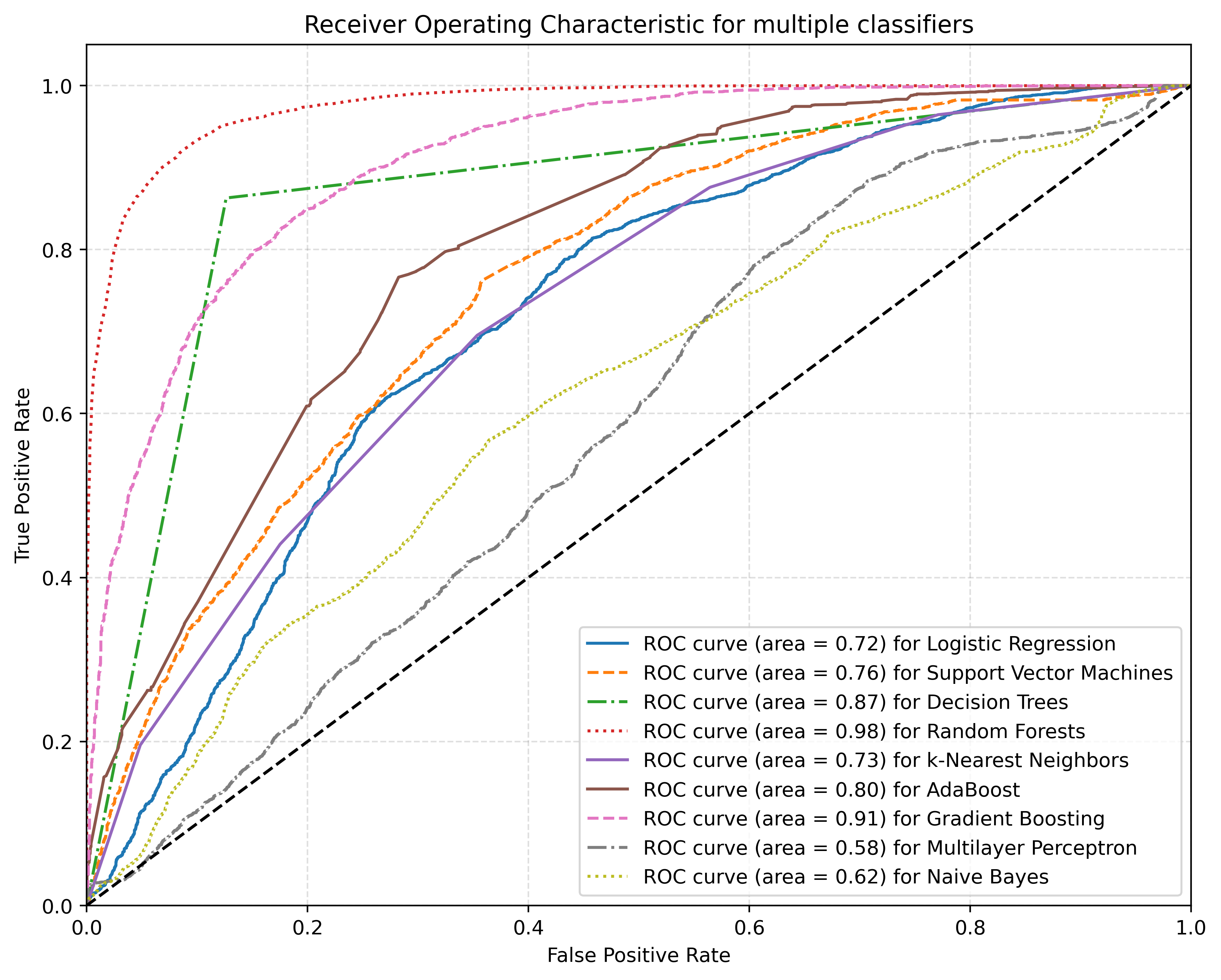}} & 
        \adjustbox{valign=c}{%
        \resizebox{6cm}{!}{%
        \begin{tabular}{
            l
            S[table-format=1.4]
            S[table-format=1.4]
            S[table-format=1.4]
            S[table-format=1.4]
        }
        \hline
        \rowcolor{mygray}
        \textbf{Classifier} & \textbf{Accuracy} & \textbf{Precision} & \textbf{Recall} & \textbf{F1-score} \\
        \hline
        Logistic Regression & 0.6248 & 0.6000 & 0.7530 & 0.6636 \\
        \rowcolor{mygray}
        Support Vector Machines & 0.6130 & 0.6304 & 0.6127 & 0.6134 \\
        Decision Trees & 0.7155 & 0.7170 & 0.7199 & 0.7229 \\
        \rowcolor{mygray}
        \textbf{Random Forests} & \textbf{0.7832} & \textbf{0.7643} & \textbf{0.8225} & \textbf{0.7911} \\
        k-Nearest Neighbors & 0.6027 & 0.6001 & 0.6521 & 0.6233 \\
        \rowcolor{mygray}
        AdaBoost & 0.6605 & 0.6549 & 0.7452 & 0.6839 \\
        Gradient Boosting & 0.7131 & 0.6918 & 0.8038 & 0.7366 \\
        \rowcolor{mygray}
        Multilayer Perceptron & 0.5949 & 0.5659 & 0.8591 & 0.5991 \\
        Naive Bayes & 0.5448 & 0.5281 & 0.8939 & 0.6620 \\
        \hline
        \end{tabular}%
    }}\\
    \hline
    \makecell[l]{\textbf{C}}  & 
    \adjustbox{valign=c}{\includegraphics[width=6.5cm]{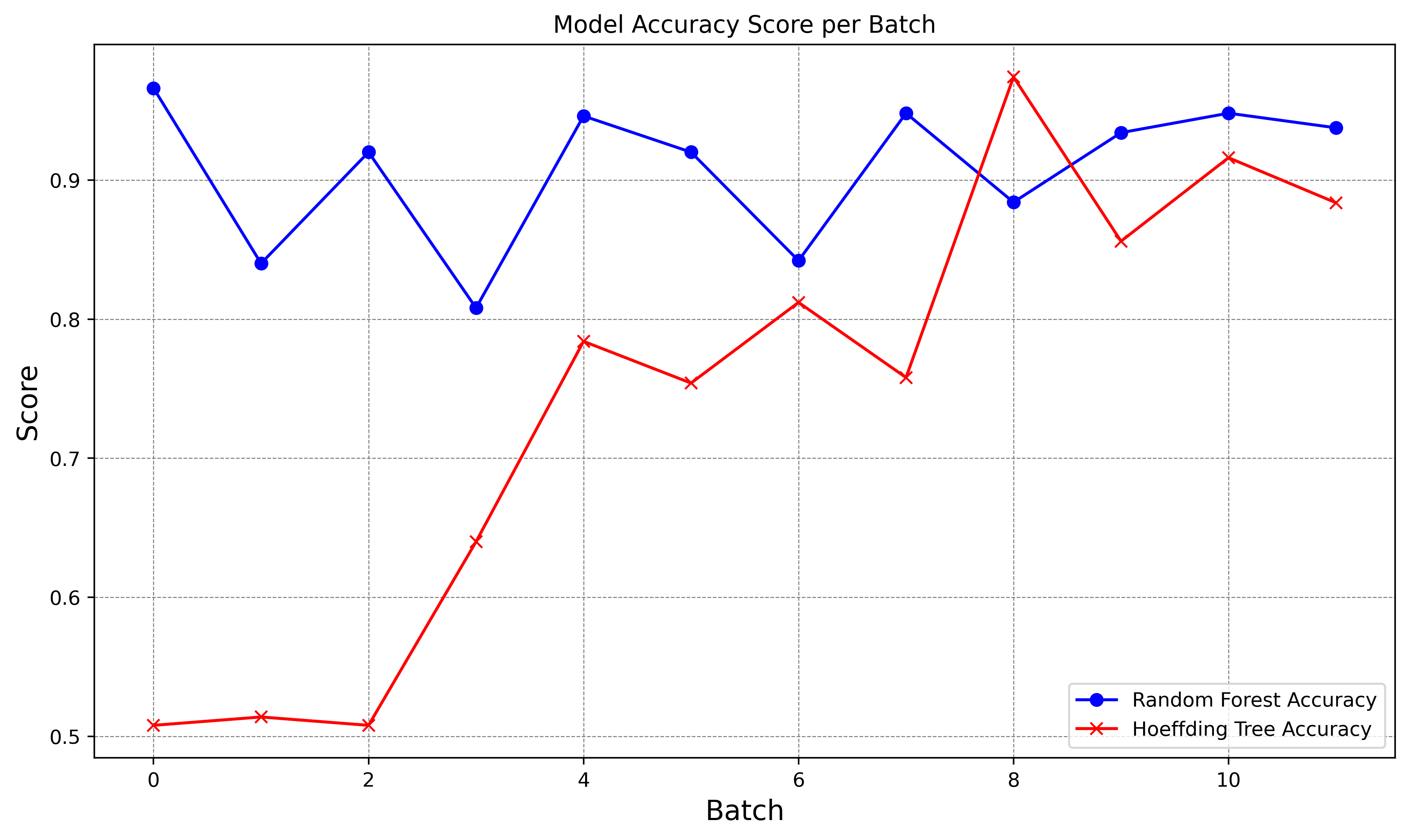}} &
    \adjustbox{valign=c}{%
        \resizebox{5.5cm}{!}{%
        \begin{tabular}{
            c
            S[table-format=1.4]
            S[table-format=1.4]
        }
        \hline
        {Batch} & {Random Forest Score} & {Hoeffding Tree Score} \\
        \hline
        \rowcolor{mygray}
        1  & 0.9660   & 0.5080       \\
        2  & 0.8400    & 0.5140       \\
        \rowcolor{mygray}
        3  & 0.9200    & 0.5080       \\
        4  & 0.8080   & 0.6400        \\
        \rowcolor{mygray}
        5  & 0.9460   & 0.7840       \\
        6  & 0.9200    & 0.7540       \\
        \rowcolor{mygray}
        7  & 0.8420   & 0.8120       \\
        8  & 0.9480   & 0.7580       \\
        \rowcolor{mygray}
        9  & 0.8840   & 0.9740       \\
        10 & 0.9340   & 0.8560       \\
        \rowcolor{mygray}
        11 & 0.9480   & 0.9160       \\
        12 & 0.9375  & 0.8836  \\
        \hline
        \end{tabular}%
    }}\\
    \hline 
    \end{tabular}
    
\end{table*}

\begin{figure*}[htbp]
    \centering
    \begin{subfigure}{0.30\textwidth}
        \includegraphics[width=\linewidth, height=4cm]{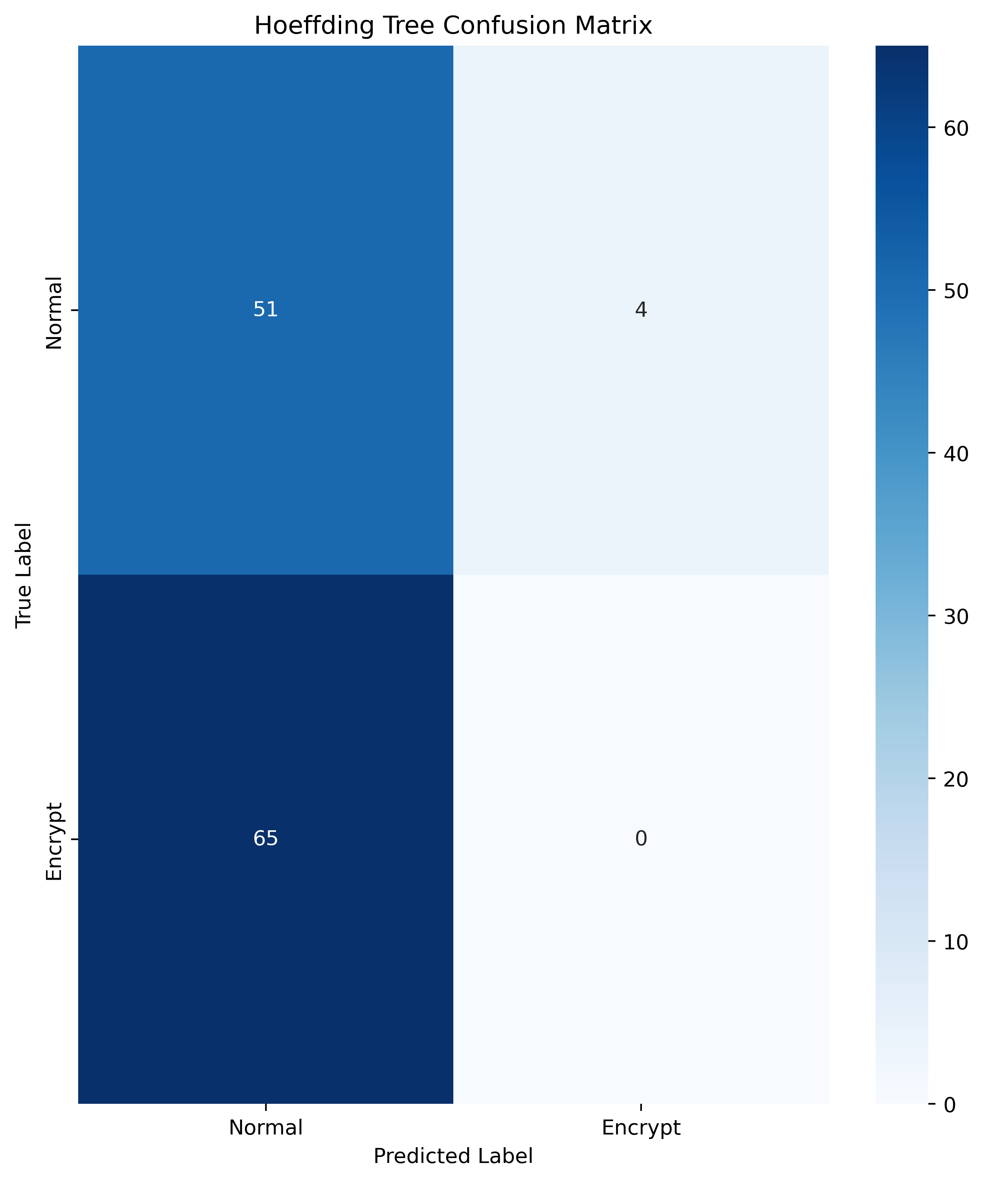}
        \caption{Batch 0: AC=0.42, (Precision Normal=0.44, Encrypt=0.0), (Recall Normal=0.93, Encrypt=0.0)}
    \end{subfigure}
    \hfill 
    \begin{subfigure}{0.30\textwidth}
        \includegraphics[width=\linewidth, height=4cm]{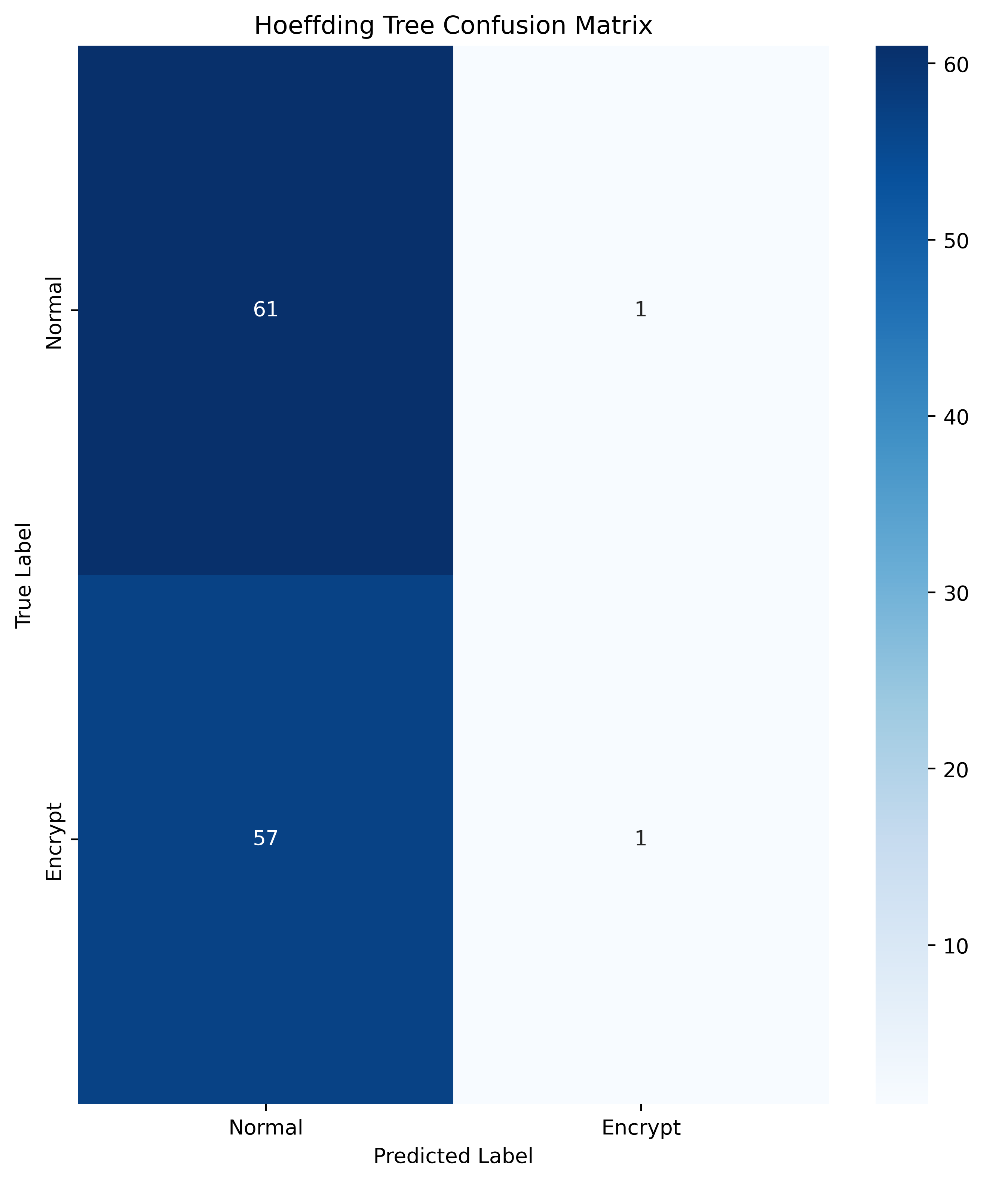}
        \caption{Batch 1: AC=0.52, (Precision Normal=0.52, Encrypt=0.50) , (Recall Normal=0.98, Encrypt=0.2)}
    \end{subfigure}
    \hfill
    \begin{subfigure}{0.30\textwidth}
        \includegraphics[width=\linewidth, height=4cm]{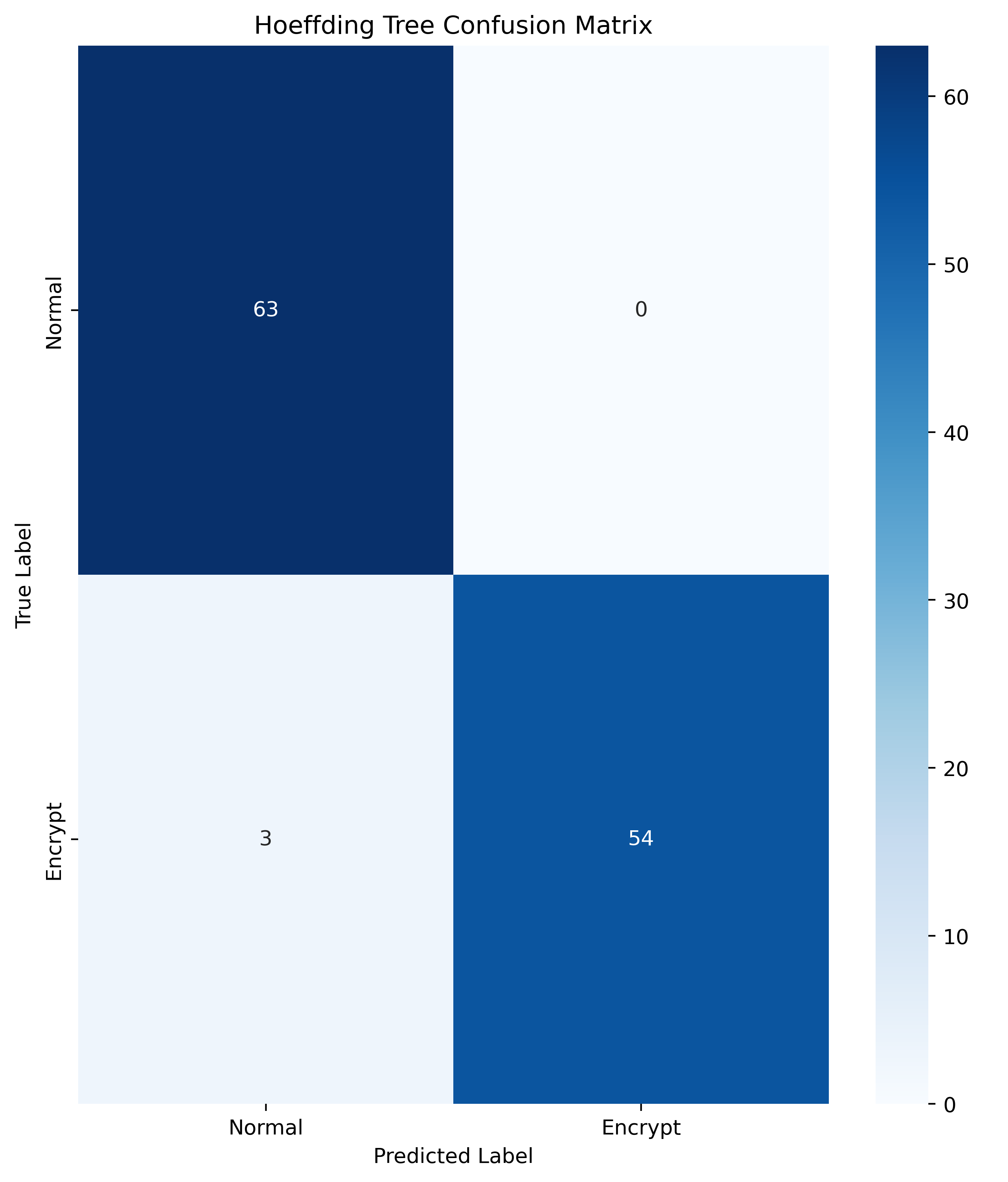}
        \caption{Batch 2: AC=0.97, (Precision Normal=0.95, Encrypt=1.0) , (Recall Normal=1.0, Encrypt=0.95)}
    \end{subfigure}
        \hfill
    \caption{The results of the Hoeffding Tree's performance on a dataset encrypted by CryptoFile ransomware involved analyzing the confusion matrix, precision, and recall across the first three batches.} \label{fig: MC-1-3}
   \end{figure*}

\begin{figure*}[htbp]
    \centering
    \begin{subfigure}{0.30\textwidth}
        \includegraphics[width=\linewidth, height=4cm]{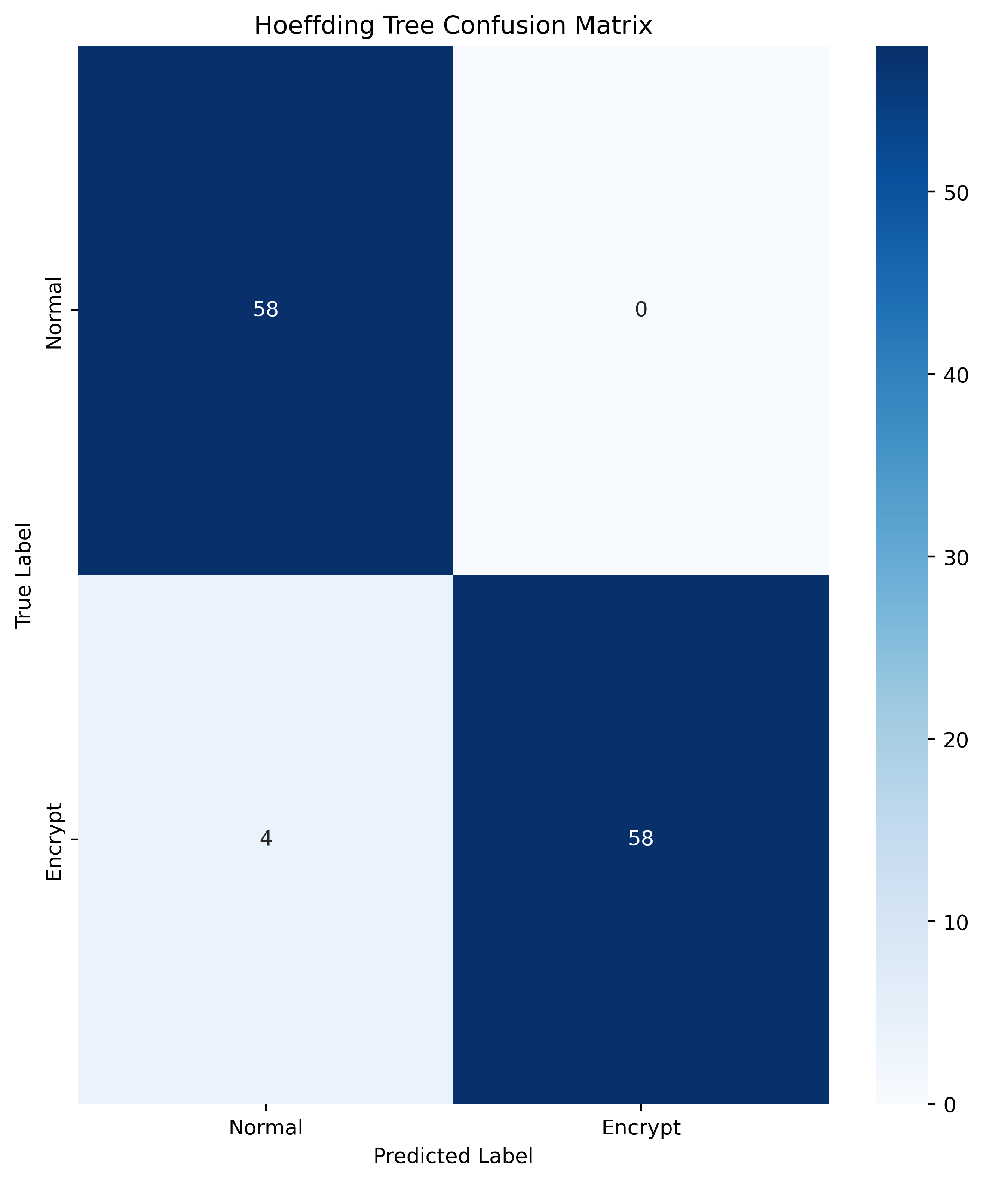}
        \caption{Batch 3: AC=0.97, (Precision Normal=0.94, Encrypt=1.0) , (Recall Normal=1.00, Encrypt=0.94)}
    \end{subfigure}
    \hfill
    \begin{subfigure}{0.30\textwidth}
        \includegraphics[width=\linewidth, height=4cm]{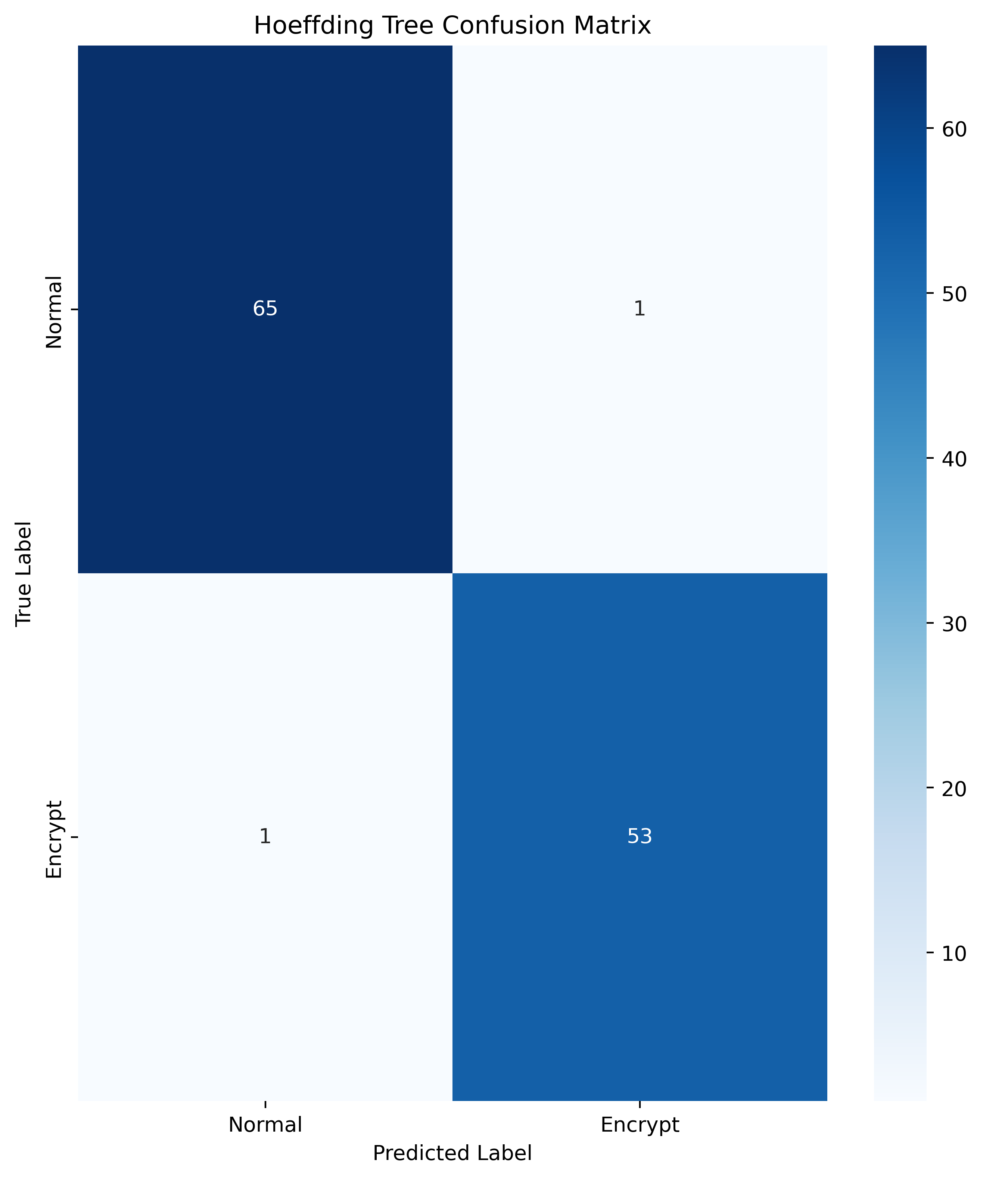}
        \caption{Batch 4: AC=0.98, (Precision Normal=0.98, Encrypt=0.98) , (Recall Normal=0.98, Encrypt=0.98)}
    \end{subfigure}
    \hfill
    \begin{subfigure}{0.30\textwidth}
        \includegraphics[width=\linewidth, height=4cm]{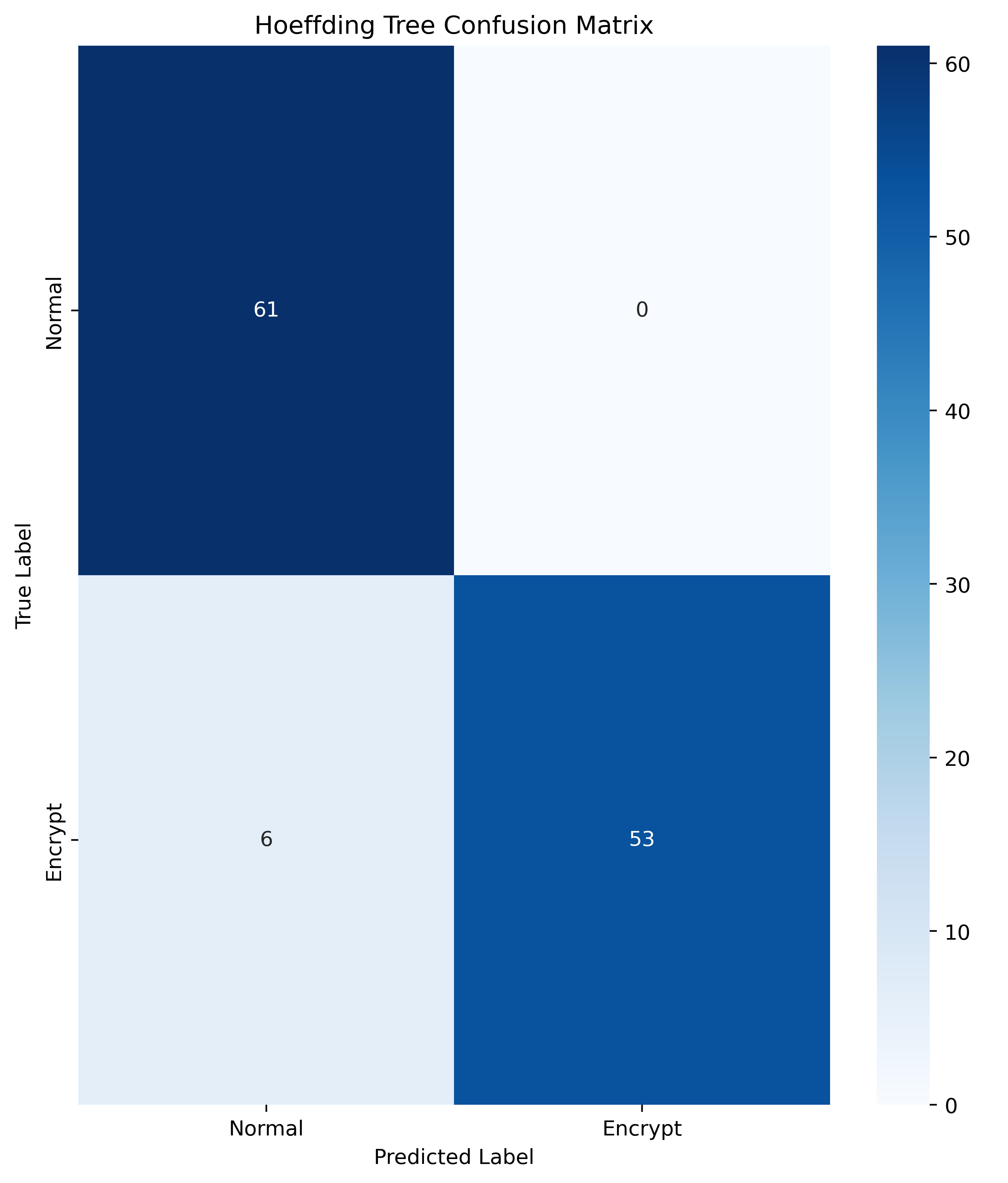}
        \caption{Batch 5: AC=0.95, (Precision Normal=0.91, Encrypt=1.00) , (Recall Normal=1.00, Encrypt=0.90)}
    \end{subfigure}
      \hfill
    \begin{subfigure}{0.30\textwidth}
        \includegraphics[width=\linewidth, height=4cm]{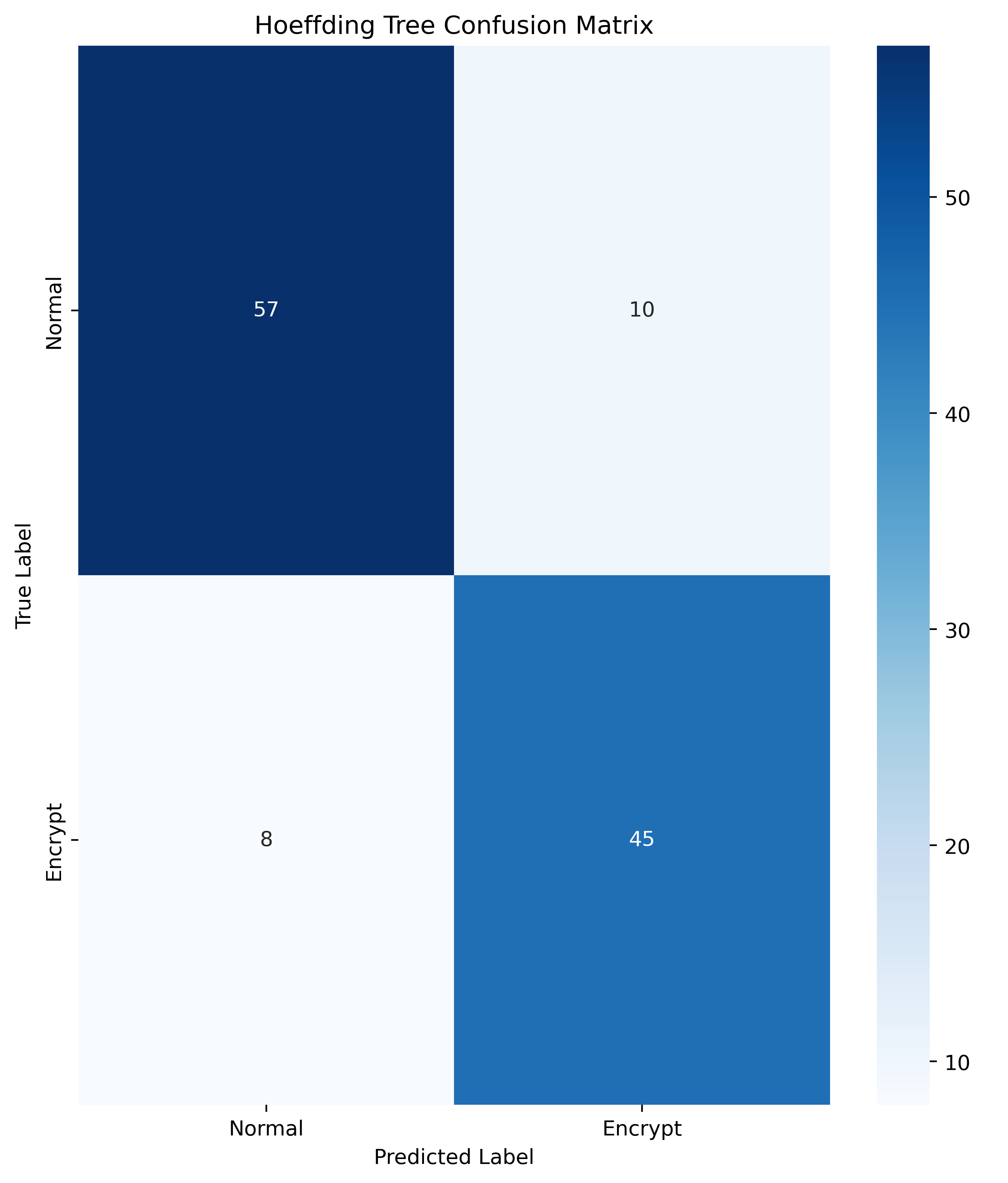}
        \caption{Batch 6: AC=0.85, (Precision Normal=0.88, Encrypt=0.82) , (Recall Normal=0.85, Encrypt=0.85)}
    \end{subfigure}
    \hfill
    \begin{subfigure}{0.30\textwidth}
        \includegraphics[width=\linewidth, height=4cm]{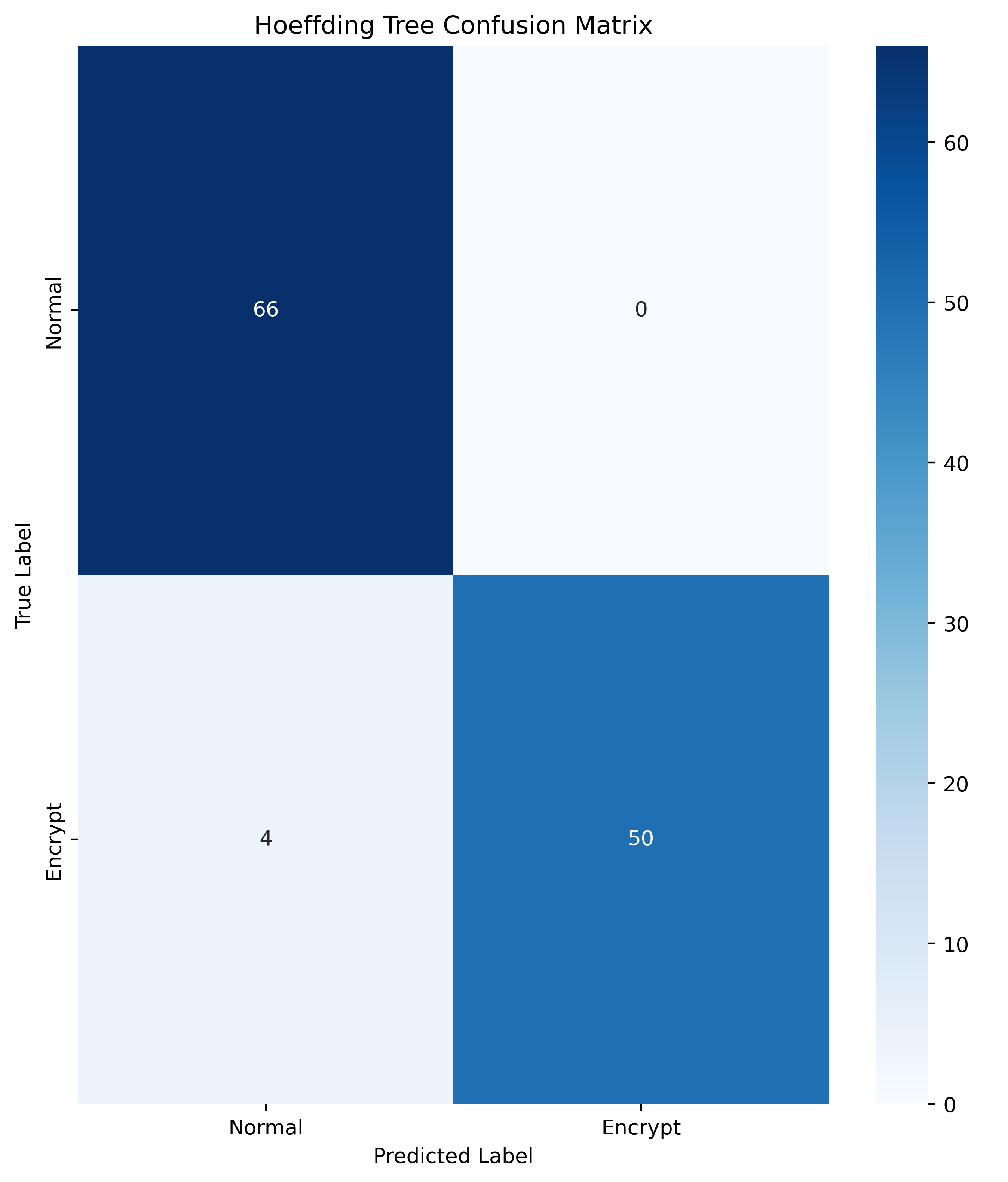}
        \caption{Batch 7: AC=0.97, (Precision Normal=0.94, Encrypt=1.00) , (Recall Normal=1.00, Encrypt=0.93)}
    \end{subfigure}
    \hfill
    \begin{subfigure}{0.30\textwidth}
        \includegraphics[width=\linewidth, height=4cm]{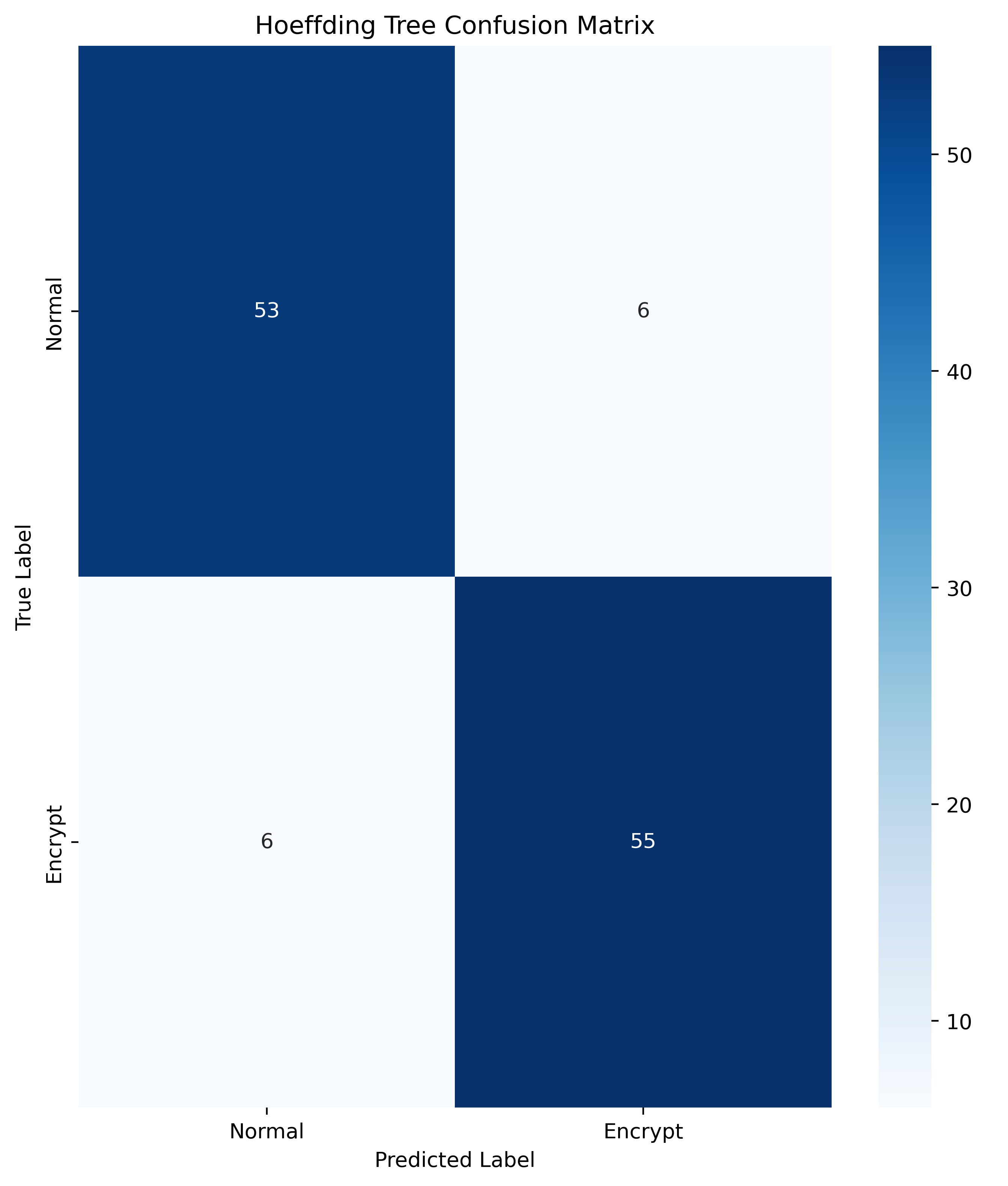}
        \caption{Batch 8: AC=0.90, (Precision Normal=0.90, Encrypt=0.90) , (Recall Normal=0.90, Encrypt=0.90)}
    \end{subfigure}
    \hfill
    \begin{subfigure}{0.30\textwidth}
        \includegraphics[width=\linewidth, height=4cm]{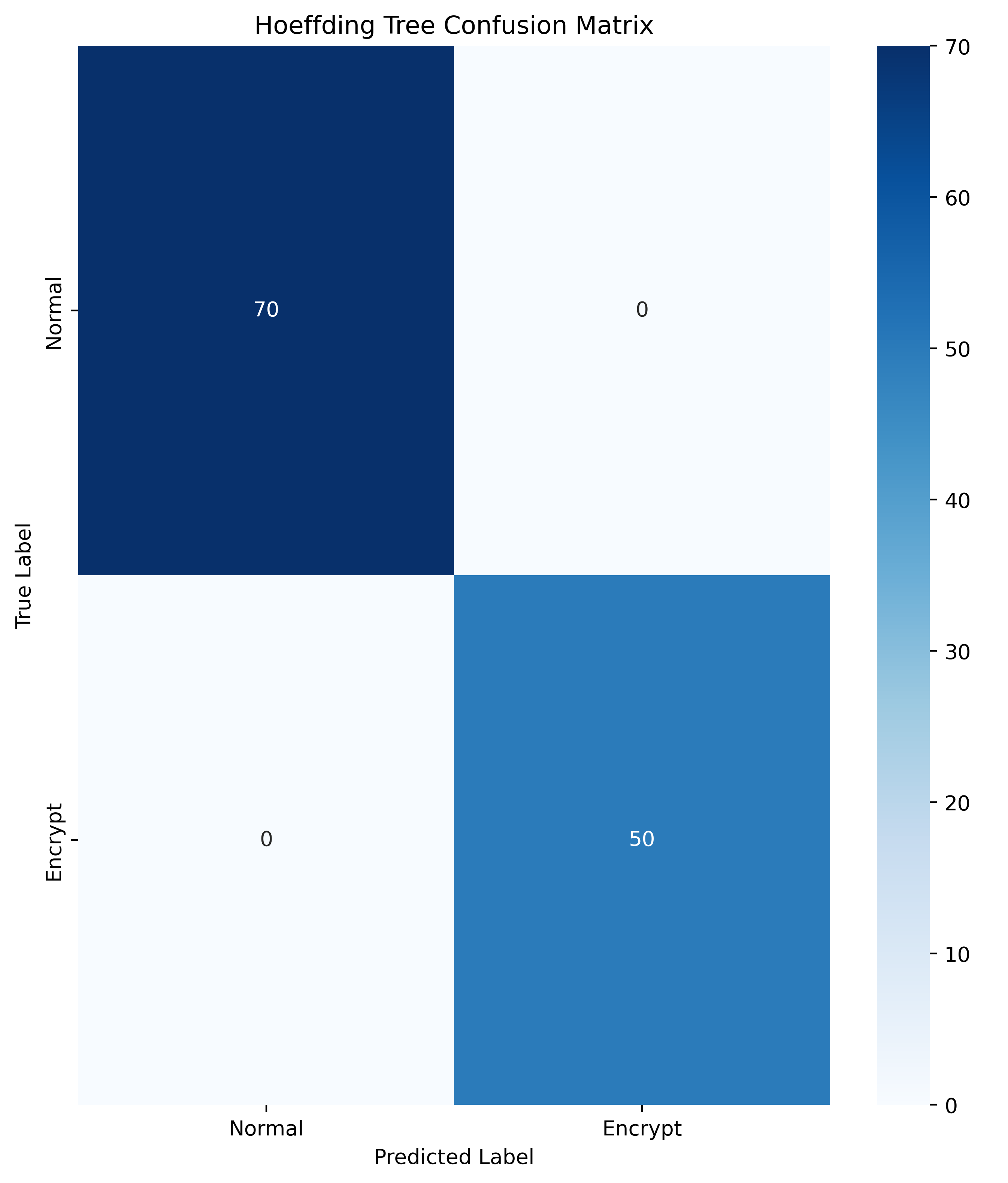}
        \caption{Batch 9: AC=1.00, (Precision Normal=1.00, Encrypt=1.00) , (Recall Normal=1.00, Encrypt=1.00)}
    \end{subfigure}
    \hfill
    \begin{subfigure}{0.30\textwidth}
        \includegraphics[width=\linewidth, height=4cm]{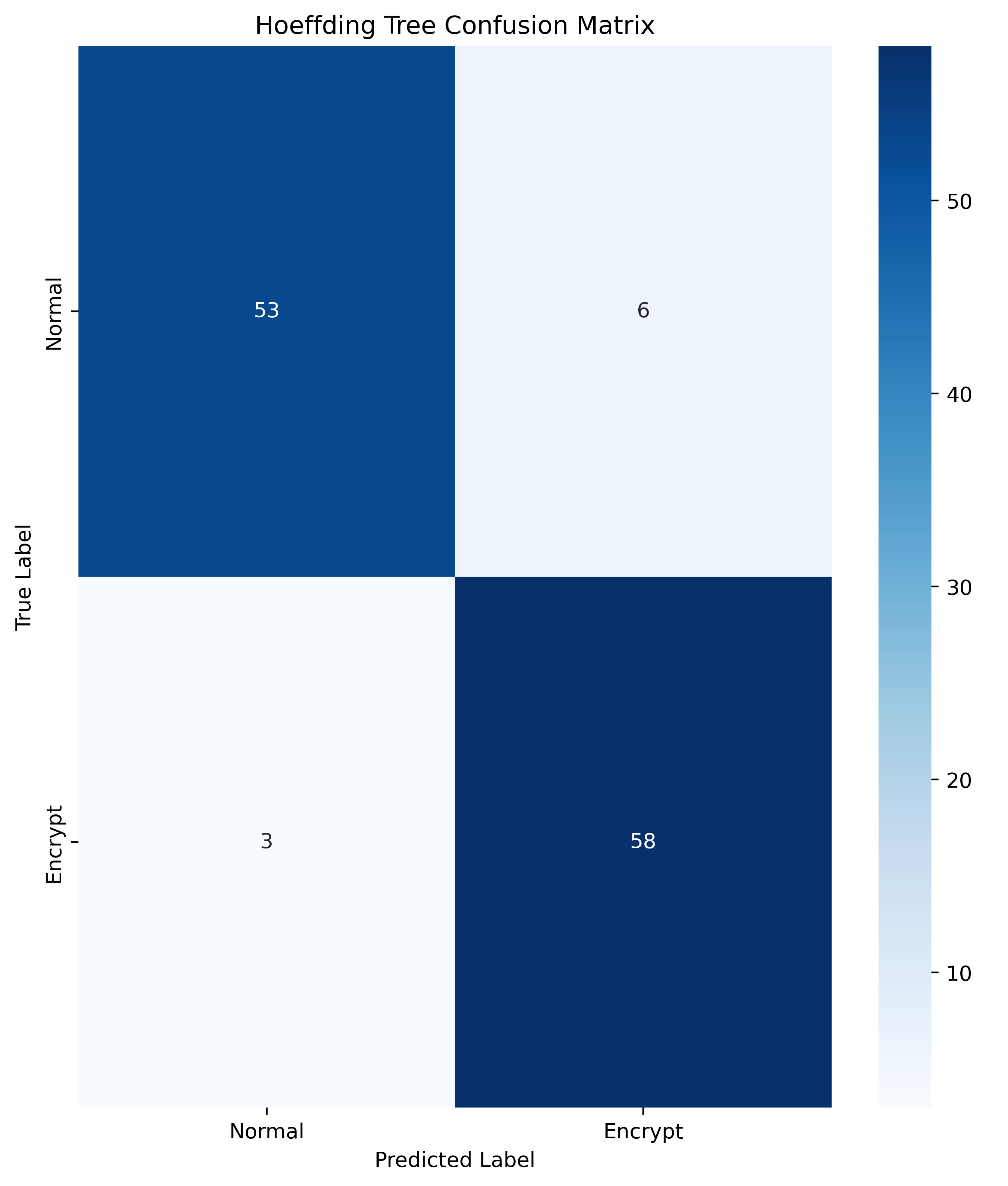}
        \caption{Batch 10: AC=0.93, (Precision Normal=0.95, Encrypt=0.91) , (Recall Normal=0.90, Encrypt=0.95)}
    \end{subfigure}
    \hfill
    \begin{subfigure}{0.30\textwidth}
        \includegraphics[width=\linewidth, height=4cm]{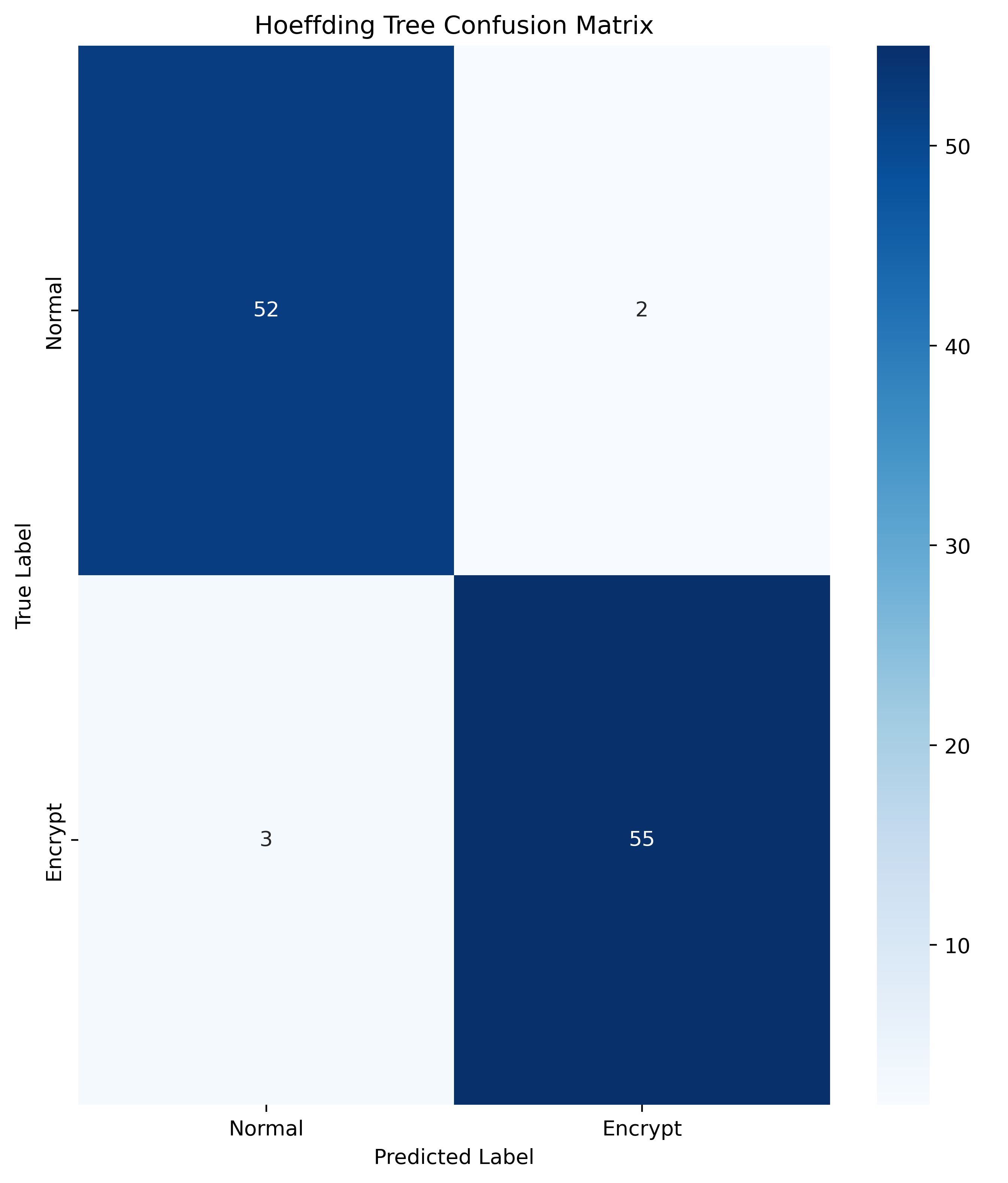}
        \caption{Batch 11: AC=0.96, (Precision Normal=0.95, Encrypt=0.96) , (Recall Normal=0.96, Encrypt=0.95)}
    \end{subfigure}
     \caption{The analysis of the Hoeffding Tree's performance on a dataset encrypted by CryptFile2 ransomware extended to include the confusion matrix, precision, and recall measurements across an additional nine batches, following the initial three illustrated in Figure \ref{fig: MC-1-3}. } \label{fig: MC-4-11}
\end{figure*}

\paragraph{\textbf{Confusion Matrix Based on Hoeffding Tree Classifier (CryptFile2 Ransomware):}}
We evaluated the model's performance using the confusion matrix and additional performance metrics like accuracy, precision, and recall. Illustrated in Figures \ref{fig: MC-1-3} and \ref{fig: MC-4-11}. The depicted figures detail a comprehensive evaluation of the Hoeffding Tree algorithm's performance on datasets affected by the CryptFile2 ransomware, focusing on the initial twelve batches of data. This analysis, segmented into two main figures, underscores the algorithm's ability to classify data as either 'Normal' or 'Encrypted' based on precision and recall metrics derived from confusion matrices.

The first figure \ref{fig: MC-1-3}, spans batches 0 through 2 and highlights a progressive improvement in the algorithm's accuracy (AC), precision, and recall rates for both normal and encrypted data classifications. Specifically, Batch 0 starts with a relatively low accuracy of 0.42, exhibiting no precision or recall for encrypted data. This trend sees a significant uptick in Batch 2, where the accuracy surges to 0.97, coupled with perfect precision and recall for encrypted data and near-perfect metrics for normal data.

The second figure \ref{fig: MC-4-11}, encompasses batches 3 through 11, providing a broader view of the algorithm's performance over time. Notably, batches 3 and 4 maintain high accuracy levels of 0.97 and 0.98, respectively, with Batch 9 achieving a perfect score of 1.00 across all metrics. This extended analysis underscores a notable consistency in the Hoeffding Tree's ability to accurately differentiate between normal and encrypted data, with a slight dip in Batch 6 before resuming high performance in subsequent batches.

\paragraph{\textbf{The intermittent encryption and online learning Challenges:}}
Given the pronounced advantages conferred to adversarial entities and the feasibility of its deployment, it is imperative to concentrate on the exploration of intermittent encryption, which constitutes a minor fraction of files encrypted by ransomware. Consequently, we posit that contemporary online learning paradigms may not adequately adapt to the nuances of intermittent encryption. In this section, we subject our dataset to encryption via the Black Basta algorithm. Table \ref{tab:Comparetive Table 2} delineates the performance metrics of various online learning classifiers, namely SGD, Perceptron, Passive Aggressive, and Hoeffding Tree algorithms, spanned across 12 distinct batches. It is salient to note that the Hoeffding Tree algorithm manifests favorable outcomes in merely four of the delineated batches.

\paragraph{\textbf{Shortcomings of the Hoeffding Tree algorithm in intermittent encryption:}} we advocate for the integration of differential entropy  \cite{michalowicz2013handbook} as a pivotal feature in the classification process of intermittently encrypted data. The differential entropy allows for a flexible and adaptive measure of the data's uncertainty, making it more suitable for dealing with the fluctuations and gaps caused by intermittent encryption.

The differential entropy, represented as \( h(X) \), serves as a potent tool in quantifying the uncertainty of continuous random variables, articulated mathematically as:

\begin{equation} 
h(X) = - \int_{-\infty}^{\infty} f(x) \log_2 f(x) \,dx
\end{equation}
where \( f(x) \) signifies the probability density function (pdf) of the variable \( X \).

By employing differential entropy, we aim to harness its analytical acumen to navigate the encrypted buffers' intricacies within the file system adeptly, thereby fostering a model imbued with enhanced precision and reliability in the face of intermittent encryption's unpredictabilities.

Conversely, a more in-depth exploration reveals that the Random Forest classifier notably excels, manifesting predominant efficacy in predicting outcomes under the influence of intermittent encryption (See results in Table \ref{tab:Comparetive Table 2}).

\subsection{Baseline model and discussion}
Recent literature emphasizes that the majority of ransomware detection mechanisms proposed for detecting encryption activities at the file system level predominantly utilize entropy analysis \cite{may2019combating, 8772046, McIntosh, s24051446, kim2022byte, vonderassen2024guardfs, 8939214}. As delineated in Section \ref{sec: Add_game}, adversaries have evolved their strategies to either circumvent entropy detection measures or reduce the visibility of encryption indicators, thereby undermining the effectiveness of conventional machine learning models. These models, which largely depend on static offline training, are increasingly ineffectual against the sophisticated techniques employed by modern adversaries. As part of our baseline evaluation, we assessed the performance of popular traditional classifiers such as SVM, Decision Trees, Random Forests, etc., all of which leverage Shannon entropy as a key feature. For this purpose, we utilized the Black Basta ransomware to encrypt a dataset, subsequently comparing the outcomes with our initial evaluation that employed the same traditional classifiers but augmented with a set of identified features, as discussed in Section \ref{sec:feature}. The findings from this evaluation are encapsulated in Table \ref{tab:classification_reports_onlyEntropy}.

In the comparative analysis of classification algorithms based on their performance metrics—accuracy, precision, recall, and F1 score—it is evident that the inclusion of a multifaceted feature set encompassing byte entropy, byte variance, byte mean, byte kurtosis, byte skewness, differential entropy, and Shannon entropy, as opposed to relying solely on entropy, significantly enhances classifier performance. Table \ref{tab:classification_reports_onlyEntropy}, characterized by its singular reliance on entropy as a feature, yielded modest performance metrics across various classifiers, with Decision Trees achieving the highest accuracy of 0.5331 and k-Nearest Neighbors demonstrating the best recall and F1 score, at 0.6710 and 0.5750, respectively. This suggests a limited capability of entropy alone to capture the intricate patterns within the data, thus constraining the classifiers' effectiveness.

Conversely, the integration of a comprehensive feature set in the second dataset led to a remarkable improvement in all evaluated metrics (see table \ref{tab:classifier_performance11}). Notably, Decision Trees exhibited a profound increase in accuracy to 0.9808, while Random Forests achieved superior precision and recall rates of 0.9703 and 0.9947, respectively. Such improvements underscore the critical importance of feature diversity in enhancing the predictive accuracy and reliability of classifiers. The multifaceted feature set evidently provides a richer, more nuanced representation of the data, thereby enabling more complex classifiers like Random Forests, Gradient Boosting, and AdaBoost to exploit this complexity to achieve near-optimal performance.

\begin{table}[h!] 
\centering

\caption{Classification reports summary for various models that rely on entropy feature encrypted by Black Basta (intermittent encryption) and CryptFile2 ransomware that encrypted by a strong encryption with RSA-2048.}\label{tab:classification_reports_onlyEntropy}

\resizebox{8cm}{!}{%
\begin{tabular}{
  l
  S[table-format=1.4]
  S[table-format=1.4]
  S[table-format=1.4]
  S[table-format=1.4]
}
\toprule
\rowcolor{mygray}
\textbf{Classifier} & {\textbf{Accuracy}} & {\textbf{Precision}} & {\textbf{Recall}} & {\textbf{F1}} \\
\midrule
Logistic Regression & 0.4881 & 0.4878 & 0.3807 & 0.4165 \\
\rowcolor{mygray}
SVM & 0.5018 & 0.5019 & 0.4006 & 0.4154 \\
Decision Trees & 0.5331 & 0.5390 & 0.4618 & 0.4948 \\
\rowcolor{mygray}
Random Forests & 0.5298 & 0.5331 & 0.4877 & 0.5068 \\
k-Nearest Neighbors & 0.5062 & 0.5064 & 0.6710 & 0.5750 \\
\rowcolor{mygray}
AdaBoost & 0.5012 & 0.4999 & 0.5952 & 0.5309 \\
Gradient Boosting & 0.5128 & 0.5110 & 0.5714 & 0.5339 \\
\rowcolor{mygray}
Multilayer Perceptron & 0.4875 & 0.4863 & 0.3769 & 0.4135 \\
Naive Bayes & 0.4940 & 0.4983 & 0.4804 & 0.4691\\
\bottomrule
\end{tabular}
}
\end{table}

\section{Related work} \label{Sec: Related_Work}

Over recent years, the field of ransomware detection has received significant attention within the academic community. Traditional detection methodologies typically involve the analysis of malicious executable \cite{zhang2020ransomware}, an approach that is increasingly proving to be ineffective against the mounting sophistication of polymorphism and code obfuscation techniques employed by malicious actors \cite{moser2007limits}.

In a concerted effort to overcome these increasingly complex obfuscation tactics, academic research has gradually transitioned toward behavior-based analysis techniques. These methods revolve around the scrutiny of runtime behaviors, such as system call sequences and Read/Write operations, which are intrinsically challenging to manipulate without fundamentally altering the malware's core functionality \cite{tian2010differentiating, kharaz2016unveil}. For instance, in the context of process-level monitoring, any ongoing process that exceeds a predefined trust threshold is identified and classified as a potentially malicious activity.  However, it is important to note that these behavior-based detection techniques are not without their limitations. Indeed, sophisticated adversaries may circumvent these strategies by dividing the ransomware activity into multiple distinct processes. Each process performs only a fraction of the overall ransomware operations, thereby resulting in the same net output as a full-scale ransomware execution \cite{de2020naked}. A prominent example of a ransomware family employing this evasion technique is LockerGoga \cite{marcus2010mcafee}.

The aforementioned behavior-based detection methodologies generally demonstrate efficacy in dealing with known threats, yet their effectiveness diminishes in the face of previously unseen ransomware samples. Given the propensity for ransomware to attack user files, recent scholarship has advocated for the implementation of a decoy file-based deceptive solution as a means to expediently detect cryptographic ransomware attacks. These decoy files, which can be created either automatically or manually, are strategically dispersed across the network \cite{whitham2017automating}.

Another direction within the field of ransomware detection involves leveraging the communication between the Command and Control (C\&C) Server and the targeted victim as a behavioral characteristic to detect potential ransomware attacks \cite{cabaj2016using, alhawi2018leveraging, almashhadani2019multi}. Almashhadani et al. \cite{almashhadani2019multi} empirically demonstrated this approach, utilizing network traffic data to investigate Locky Ransomware. In their study, they extracted 20 features from the network traffic data to construct a classification model. Experimental evaluations of this proposed detection system using Random Forests, Bayes Networks, and Support Vector Machines indicated high detection accuracy, a low false positive rate, and useful feature extraction. Additionally, the system proved highly effective in tracking the network activities of ransomware. However, there are important caveats to consider in the application of these techniques. Firstly, Almashhadani et al.'s work focuses solely on the Locky ransomware, which introduces potential issues of generalisability, as the findings might not apply to other types of ransomware. Secondly, while some ransomware families necessitate an internet connection to initiate the encryption process, numerous others do not require a connection to the C\&C server to begin encrypting victim files, thereby limiting the effectiveness of this approach.

A different behavioral feature used for ransomware detection is the monitoring of hardware performance counters (HPCs) such as cache-reference, cache misses, and instruction counts \cite{alam2019ratafia, aurangzeb2021classification}. This methodology is rooted in the fact that ransomware perpetrators, in their quest to maximize potential gains, seek to encrypt as many user files as possible. The aggressive execution of file encryption and renaming processes usually provokes a context switch, which in turn influences the CPU status, affecting elements such as the CPU cache and branch prediction. For instance, the authors in EGB \cite{aurangzeb2021classification} evaluated the features derived from hardware performance counters using several machine learning algorithms to classify applications into ransomware and non-ransomware categories.  In their research, EGB \cite{aurangzeb2021classification} demonstrated that a model built using the Random Forest algorithm exhibited the highest detection rate.

Additionally, DeepGuard's work \cite{ganfure2020deepguard} introduced an innovative method for capturing user activity. Their proposed mechanism relied on monitoring file interaction patterns and utilized a deep generative autoencoder architecture to reconstruct the input. DeepGuard's approach is premised on the assumption that, with sufficient training on user file-interaction logs collected over several days, the deep generative autoencoder becomes proficient in reconstructing unseen inputs with minimal reconstruction error, provided the input resembles legitimate user activity. Consequently, this model can identify anomalous (or ransomware) activity by exclusively modeling the user interaction pattern, effectively distinguishing benign user activities from ransomware operations.

A study by Baek et al. \cite{baek2020ssd} proposed a two-pronged ransomware protection system known as SSDinsider++. This system focused both on the detection of ransomware attacks and on data recovery. However, their approach was constrained by the practical infeasibility of assessing every I/O block, its header, and payload during runtime. SSDinsider++'s detection algorithm operates by observing the I/O patterns of a host system and determines whether the host is under ransomware attack at an early stage.

Meanwhile, in an effort to enhance the detection rate of crypto-ransomware, the authors in File-entropy \cite{hsu2021enhancing} leveraged the entropy of encrypted files as a metric to identify ransomware activity. They encrypted and analyzed over 20 file formats using WannaCry, Phobos, GandCrab, and Globelmposter ransomware, gathering features essential for distinguishing ransomware tasks from file encryption activities, such as Zip and 7z. These extracted features were then fed into a support vector machine for classification tasks. Their experimental results showed that their model could detect ransomware activity with an average accuracy of 85.17\%. However, their research was limited to only four ransomware samples, and may therefore lack the ability to detect other varieties of ransomware.

The deception-based \cite{10026355} solution operates by setting up artificial computer system assets such as a web server or router across an enterprise network. These assets act as bait, enticing intruders and subsequently triggering alerts when accessed. While this methodology is still relatively novel, some recent studies have proposed deception-based strategies for ransomware detection \cite{lee2017make, GOMEZHERNANDEZ2018389, 10026355, mehnaz2018rwguard}. 

For example, Lee et al. \cite{lee2017make} implemented a proof of concept involving the creation and placement of decoy files based on ransomware file traversal patterns. In their study, various ransomware samples were decompiled and analyzed to understand how ransomware traverses file systems. They suggested planting two deceptive files in the root directory, one at the beginning and another at the end of an alphabetical list. Additionally, Gómez et al. \cite{GOMEZHERNANDEZ2018389} developed R-Locker, a tool designed to thwart ransomware attacks by creating decoy files. These files were generated using the 'makeinfo' command in Unix systems, filled with 3KB of data, and placed in the user's home directory. R-Locker would then block any process attempting to access these files. However, the effectiveness of this approach was limited because all the decoy files were of the same type, allowing ransomware to potentially evade detection. 

In Rwguard \cite{mehnaz2018rwguard}, the authors proposed a novel combination of a behavioral-based detection model with a decoy-based approach to combat cryptographic ransomware attacks. Rwguard randomly designated original user files as decoy files. Likewise, Cryptostopper \cite{10026355} embedded arbitrarily generated decoy files in the file system to create a deceptive defense against ransomware attacks. Any attempt to alter the content of these deceptive files would trigger an alert, notifying the system administrator and closing the infected host. As Cryptostopper is a commercial product, its technical details are not publicly available.

Several studies have utilized dynamic analysis to examine the behavior of ransomware, including work by Kharraz et al., Song et al., Cabaj et al., and Andronio et al. \cite{cabaj2015network, kharaz2016unveil, mbol2016efficient, song2016effective, andronio2015heldroid}. These studies focused on various elements such as I/O data buffer entropy, access patterns, file system activities \cite{Mahboubi10}, CPU and memory usage, and network behavior. Despite its effectiveness against evasive ransomware types, dynamic analysis has inherent limitations. Notably, it necessitates the execution of ransomware in a safe and monitored environment; otherwise, the platform risks infection. This constraint may limit the feasibility of dynamic analysis in certain scenarios. We have summarized and critiqued related work approaches, as well as what this paper offers, in Table \ref{tab:summary}.

\begin{table*}[h!]
\centering
    \caption{Summary and Critique of Ransomware Detection Approaches}
    \label{tab:summary}
    \footnotesize
    \begin{tabular}{|p{3cm}|p{5cm}|p{7cm}|}
        \hline
        \textbf{Approach} & \textbf{Criteria} & \textbf{Summary and Critique} \\
        \hline
        Behaviour-Based Analysis & - Effectiveness against polymorphism and obfuscation techniques \cite{moser2007limits} & Effective against evolving ransomware tactics, but less effective against unknown ransomware \cite{tian2010differentiating}. \\
                                & - Efficacy against known threats \cite{zhang2020ransomware} & Demonstrates high efficacy. \\
        \hline
        Decoy File-Based Deception & - Detection speed and accuracy & Rapidly detects cryptographic ransomware attacks. \\
                                & - Lack of deception diversity & Limited effectiveness when ransomware adapts to recognize decoy files as fake. \\
        \hline
        Communication Analysis & - Detection accuracy and false positive rate \cite{almashhadani2019multi} & High accuracy and low false positive rate. \\
                               & - Limited applicability & Limited effectiveness when ransomware doesn't rely on C\&C servers or when dealing with new, previously unseen variants. \\
        \hline
        Hardware Performance Counters & - Detection rate \cite{aurangzeb2021classification} & High detection rate with hardware performance counters. \\
                                    & - Resource overhead \cite{alam2019ratafia} & May introduce resource overhead, impacting system performance. \\
                                    & - Limited applicability & May not be effective against non-resource-intensive ransomware. \\
        \hline
        User Activity Monitoring & - Anomaly detection capability \cite{ganfure2020deepguard} & Effective at distinguishing benign activities from ransomware. \\
                               & - Training data limitations & Requires substantial training data and may struggle with rapidly evolving ransomware. \\
        \hline
        SSDinsider++ & - Early-stage detection capability \cite{baek2020ssd} & Early detection based on I/O patterns. \\
                    & - Practicality \cite{baek2020ssd} & Practicality limited due to resource-intensive assessment. \\
        \hline
        File Entropy Analysis & - Detection accuracy \cite{hsu2021enhancing} & Accurate in detecting ransomware activity. \\
                            & - Limited ransomware sample coverage \cite{hsu2021enhancing} & Limited to the ransomware samples used for training. \\
                            & - Scalability \cite{hsu2021enhancing} & May struggle to scale with a large number of file formats. \\
        \hline
        Deception-Based Strategies & - Intruder enticement and alert triggering \cite{lee2017make} & Detects intruders and triggers alerts effectively. \\
                                & - Effectiveness against adaptive ransomware & Limited effectiveness when ransomware evolves to recognize decoy assets. \\
        \hline
        Dynamic Analysis & - Effectiveness against evasive ransomware types \cite{cabaj2015network} & Effective but requires a safe and monitored environment. \\
                        & - Platform infection risk \cite{cabaj2015network} & Necessitates executing ransomware in a controlled environment, which may not always be feasible due to virtualization detection, incubation period, and other evasion techniques \cite{KharrazAmin}. \\
                        & - Variability among ransomware variants \cite{KharrazAmin} & Behaviours are specific to individual ransomware variants, requiring generic models or continuous training to adapt to new variants. \\
        \hline
    \end{tabular}
\end{table*}

Table \ref{tab:summary1} serves as a concise overview of various ransomware detection approaches, highlighting their strengths and limitations. It offers a quick reference for researchers and practitioners in the field of cybersecurity, providing insights into the key criteria used to evaluate these approaches. The table aims to assist in understanding the relative effectiveness of each method in combating ransomware threats and what this paper is aiming to achieve. The paper examines techniques employed by adversaries to diminish the indication of compromise threshold by manipulating various features and activities, including factors like entropy, input and output data, and the exhaustive utilization of applications.

\begin{table*}[h!]
    \centering
    \caption{Summary and Critique of Ransomware Detection Approaches: In this context, \textbf{IE} stands for Intermittent Encryption, \textbf{PE} stands for Partial Encryption, and \textbf{AB} stands for AES-Base64 encryption and encoding.}
    \label{tab:summary1}
    \footnotesize 
    \begin{tabular}{|p{3cm}|p{3.5cm}|p{6cm}|p{0.4cm}|p{0.4cm}|p{0.4cm}|}
        \hline
        \textbf{Approach} & \textbf{Contributions} & \textbf{Challenges} & \textbf{IE} & \textbf{PE} & \textbf{AB} \\
        \hline
        Behaviour-Based Analysis & - Polymorphism and obfuscation techniques \cite{moser2007limits} &  - Limited effectiveness against unknown ransomware \cite{tian2010differentiating}. & \xmark & \xmark & \xmark \\
                               & - Efficacy against known threats \cite{zhang2020ransomware} & & & & \\
        \hline
        Decoy File-Based Deception & - Detection speed and accuracy & - Limited effectiveness when ransomware adapts to recognize decoy files as fake & \xmark & \xmark & \xmark \\
                               & & - Lack of deception diversity & & & \\
        \hline
        Communication Analysis & - Detection accuracy and false positive rate \cite{almashhadani2019multi} & - Limited effectiveness when ransomware does not rely on C\&C servers or when dealing with new variants & \xmark & \xmark & \xmark \\
                               & & - Limited applicability & & & \\
        \hline
        Hardware Performance Counters & - High detection rate with hardware performance counters \cite{aurangzeb2021classification} & - Resource overhead \cite{alam2019ratafia} & \xmark & \xmark & \xmark \\
                                    & & - Limited applicability & & & \\
                                    & & - Limited effectiveness against non-resource-intensive ransomware & & & \\
        \hline
        User Activity Monitoring & - Anomaly detection capability \cite{ganfure2020deepguard} & - Requires substantial training data and may struggle with rapidly evolving ransomware & \xmark & \xmark & \xmark \\
                               & - Effective at distinguishing benign activities from ransomware & & & & \\
        \hline
        SSDinsider++ & - Early-stage detection on I/O patterns capability \cite{baek2020ssd} & - Practicality limited due to resource-intensive assessment \cite{baek2020ssd} & \xmark & \xmark & \xmark \\
        \hline
        File Entropy Analysis & - High detection accuracy \cite{hsu2021enhancing} & - Limited ransomware sample coverage \cite{hsu2021enhancing} & \xmark & \xmark & \xmark \\
                            & & - Scalability \cite{hsu2021enhancing} & & & \\
        \hline
        Deception-Based Strategies & - Intruder enticement and alert triggering effectively \cite{lee2017make} & - Limited effectiveness against adaptive ransomware & \xmark & \xmark & \xmark \\
                                & & - Limited effectiveness when ransomware evolves to recognize decoy assets & & & \\
        \hline
        Dynamic Analysis & - Effectiveness against evasive ransomware types \cite{cabaj2015network} & - Platform infection risk that necessitates executing ransomware in a controlled environment, which may not always be feasible \cite{cabaj2015network} & \xmark & \xmark & \xmark \\
                        & & - Variability among ransomware variants \cite{KharrazAmin}, specifically behaviors are specific to individual ransomware variants, requiring generic models or continuous training to adapt to new variants. & & & \\
                        & & - Behaviours are specific to individual ransomware variants, requiring generic models or continuous training to adapt to new variants. & & & \\
        \hline
        This paper & - Effectiveness against encryption evasive methods, statistical analysis of content data & - Online learning involves acting as a file share proxy within the file system, such as FUSE, and constantly training on input data from a network shared drive. attack adaptability and the use of not yet known techniques   & \cmark & \cmark & \cmark \\
        \hline
    \end{tabular}
\end{table*}

\section{Conclusion and future directions} \label{sec: Conclusion}
This study delves deeply into the ever-expanding realm of ransomware, meticulously investigating the dynamic confrontations between adversaries, who continuously evolve sophisticated strategies for optimizing illicit gains, and defenders, who carefully orchestrate countermeasures to mitigate risks associated with file system-level data encryption. Through a comprehensive analysis, the research illuminates nuanced adversary techniques designed to circumvent storage-level and behavioral countermeasures, and employs advanced methodologies, like online incremental machine learning algorithms, to unravel the complexities of these confrontations. The findings reveal that the Hoeffding Tree algorithm, renowned for its intrinsic incremental learning abilities, stands out with exemplary performance, particularly against strong, partial, and Base64 encoding encryption techniques, demonstrating an impressive accuracy average above 90\%. In contrast, the Random Forest classifier, enriched with a warm-start feature set, prevails with remarkable efficacy, achieving a minimum of 80\% accuracy, specifically in scenarios involving intermittent encryption. In the forthcoming phases of our research, the vision is to innovate further by deploying a hybrid model within our developmental framework, intending to fortify the proxy gateway designed to secure endpoints connecting to cloud storage. Data preprocessing involves extracting relevant features from files, a crucial step in enabling classifiers to effectively distinguish between encrypted and unencrypted files. After a comprehensive review of all ransomware techniques outlined in Section \ref{sec: Add_game}, we identified and extracted features that can be roughly categorized into three groups: (1) Features related to file entropy; (2) Features related to file size; and (3) Feature related to file content/pattern. We also investigate how Asymmetric Numeral Systems (ANS), which provide both efficient and near-optimal entropy coding, could be leveraged by ransomware developers as part of their tactics \cite{camtepe2022ans, e25040672}.

\bibliographystyle{elsarticle-num}
\bibliography{DeltaFileGuard}

\end{document}